\documentclass[lettersize,journal]{IEEEtran}
\usepackage{amsmath,amsfonts}
\usepackage{algorithmic}
\usepackage{array}
\usepackage[caption=false,font=normalsize,labelfont=sf,textfont=sf]{subfig}
\usepackage{textcomp}
\usepackage{stfloats}
\usepackage{url}
\usepackage{verbatim}
\usepackage{graphicx}
\usepackage{cite}
\usepackage{multirow}
\usepackage{makecell}
\usepackage{booktabs}
\usepackage{tabularx}
\usepackage[caption=false,font=normalsize,labelfont=sf,textfont=sf]{subfig}
\usepackage{changepage} 
\usepackage{floatrow} \usepackage{float}
\floatstyle{plaintop}
\restylefloat{table}
\usepackage[table]{xcolor}
\usepackage{graphicx}

\usepackage[colorlinks=true,citecolor=green]{hyperref}

\usepackage[linesnumbered,ruled,vlined]{algorithm2e}

\SetCommentSty{mycommfont}

\newcommand{\symbolvec}[1]{\boldsymbol{#1}}
\newcommand{\symbolmat}[1]{\mathbf{#1}}
\newcommand{\symbolset}[1]{\mathcal{#1}}

\begin{document}

\title{A Benchmark for Multi-speaker Anonymization}

\author{Xiaoxiao Miao, \IEEEmembership{Member, IEEE},
Ruijie Tao, \IEEEmembership{Member, IEEE} 
Chang Zeng,
Xin Wang, \IEEEmembership{Member, IEEE} 
\thanks{This study is partially supported by JST, PRESTO Grant Number JPMJPR23P9, Japan, SIT-ICT Academic Discretionary Fund, and Ministry of Education, Singapore, under its Academic Research Tier 1 (R-R13-A405-0005) and its SIT's Ignition grant (STEM) (R-IE3-A405-0005).}

\thanks{Xiaoxiao Miao is with the Singapore Institute of Technology, Singapore 567739 (e-mail:xiaoxiao.miao@singaporetech.edu.sg).}
\thanks{Ruijie Tao is with National University of Singapore, Singapore 117583 (e-mail:ruijie.tao@u.nus.edu)}. 
\thanks{Chang Zeng is with the National Institute of Informatics, 2-1-2 Hitotsubashi Chiyoda-ku, Tokyo 101- 8340, Japan (e-mail:zengchang.elec@gmail.com)}
\thanks{Xin Wang is with the National Institute of Informatics, 2-1-2 Hitotsubashi Chiyoda-ku, Tokyo 101- 8340, Japan (e-mail:wangxin@nii.ac.jp).}
}

\markboth{Journal of \LaTeX\ Class Files,~Vol.~14, No.~8, August~2021}%
{Shell \MakeLowercase{\textit{et al.}}: A Sample Article Using IEEEtran.cls for IEEE Journals}

\maketitle

\begin{abstract}
Privacy-preserving voice protection approaches primarily suppress privacy-related information derived from paralinguistic attributes while preserving the linguistic content. Existing solutions focus particularly on single-speaker scenarios. However, they lack practicality for real-world applications, i.e., multi-speaker scenarios. In this paper, we present an initial attempt to provide a multi-speaker anonymization benchmark by defining the task and evaluation protocol, proposing benchmarking solutions, and discussing the privacy leakage of overlapping conversations. 
The proposed benchmark solutions are based on a cascaded system that integrates spectral-clustering-based speaker diarization and disentanglement-based speaker anonymization using a selection-based anonymizer. To improve utility, the benchmark solutions are further enhanced by two 
\textit{conversation-level} speaker vector anonymization methods. The first method minimizes the differential similarity across speaker pairs in the original and anonymized conversations, which maintains original speaker relationships in the anonymized version. The other minimizes the aggregated similarity across anonymized speakers, which achieves better differentiation between speakers.
Experiments conducted on both non-overlap simulated and real-world datasets demonstrate the effectiveness of the multi-speaker anonymization system with the proposed speaker anonymizers. Additionally, we analyzed overlapping speech regarding privacy leakage and provided potential solutions\footnote{Code and audio samples are available at \url{https://github.com/xiaoxiaomiao323/MSA}, evaluation datasets can be download from \url{https://zenodo.org/records/14249171}.}.

\end{abstract}


\section{Introduction}
\label{sec:intro}
\IEEEPARstart{S}peech data is clearly defined as personally identifiable data under the EU General Data Protection Regulation (GDPR) \cite{GDPR}. Its rich information content, apart from the spoken language itself, encompasses attributes like age, gender, emotion, identity, geographical origin, and health status. Failure to implement voice privacy protection measures and directly sharing raw audio data with social platforms or third-party companies may result in privacy leakage \cite{ergunay2015vulnerability, vestman2020voice}. In the worst-case scenario, attackers could exploit advanced generative artificial intelligence technologies to clone or manipulate the original speakers' audio for voice authentication systems for illegal purposes \cite{kinnunen2010overview, hansen2015speaker, bai2021speaker}. Another area of concern involves deducing additional paralinguistic information from the raw speech, leading to the creation of applications such as targeted advertisements \cite{Lawsuit2022} based on factors like customer age, gender, and accent.

Regulations and the social awareness of the privacy issues have seen quite a few applications of the voice privacy protection techniques. For example, broadcasting companies have been using techniques to protect the identities of witness and whistle blowers in sensitive interviews\footnote{\url{https://www.nii.ac.jp/today/103/8.html} (in Japanese)}. Similar techniques are used in medical sections to protect the privacy of the patients' speech data \cite{tayebi2024addressing}. Additionally, there are growing concerns in educational systems about protecting children's speaker identities\footnote{\href{https://www.privacyworld.blog/2024/04/singapore-issues-privacy-guidelines-for-childrens-online-safety/}{Privacy Guidelines for Children's Online Safety}} while ensuring audio quality for analysis. Another application is to anonymize the speakers' identities before publishing a speech database \cite{miao2024synvox2,anonTTS}. This preserves the privacy of the original speakers while making the data available for downstream tasks.

One mainstream solution from the academic community is to implement a user-centric voice protection solution on the raw datasets before data sharing. 
In recent years, efforts to protect voice privacy have primarily concentrated on techniques like noise addition \cite{7472729}, voice transformation \cite{qian2017voicemask}, voice conversion \cite{magarinos2017reversible,jin2009voice,qian2018hidebehind,huang2021defending}, and speech synthesis \cite{fang2019speaker}. However, researchers perform these studies in diverse settings, making them incomparable.

The VoicePrivacy Challenge (VPC) held in 2020 \cite{tomashenko2021voiceprivacy}, 2022 \cite{tomashenko2022voiceprivacy}, and 2024 \cite{tomashenko2024voiceprivacy} provided a formal definition of the speaker anonymization task, common datasets, evaluation protocols, evaluation metrics, and baseline systems, promoting the development of privacy preservation techniques for speech technology. Given an input speech waveform \emph{uttered by a single speaker}, an ideal speaker anonymization system in VPC should protect speaker identity information (privacy) while maintaining linguistic and prosodic content (stress, intonation, and rhythm) to enable various downstream tasks (utility).

The primary baseline of VPC aims to separate speaker identity information from linguistic and prosodic content, generating anonymized speech where only identity information is removed. It extracts three types of features from an original speech recording: (i) a speaker vector encoding speaker identity information \cite{snyder2018x}, (ii) content features capturing linguistic content \cite{povey2018semi}, and (iii) pitch features conveying prosodic information. To hide the speaker identity of the original speaker, a speaker anonymizer searches for several farthest speaker vectors from the original speaker vector in an external speaker vector pool, then averages randomly-selected ones as the anonymized speaker vector \cite{fang2019speaker,Srivastava2020DesignCF,srivastava2022privacy}. Anonymized speech is finally generated by synthesizing speech from original content and pitch features along with the anonymized speaker vector \cite{wang2019neural}.

Following the VPC protocol, several works have proposed improvements from various aspects. These include (i) enhancing disentanglement to prevent privacy leakage from content and prosody features \cite{shamsabadi2022differentially, mawalim2022speaker, champion2022disentangled, champion2023anonymizing}, (ii) improving the speaker anonymizer to generate natural and distinctive anonymized speaker vectors that protect speaker privacy against various attackers \cite{meyer2023prosody, miao2023language, yao2024distinctive}, (iii) modifying not only speaker identity but other privacy-related paralinguistic attributes such as age, gender, and accent to enable more flexible anonymization \cite{zhang2023voicepm, noe2023hiding}, (iv) exploring language-robust speaker anonymization that supports anonymizing unseen languages without severe language mismatch \cite{miao22_odyssey,miao2022analyzing,miao2023language}.

Although these efforts have driven forward the development of speaker anonymization techniques, all of them mainly focus on single-speaker scenarios---the input utterance is assumed to contain the voice of \emph{a single speaker}. This paper refers to this as single-speaker anonymization (SSA).

Compared with SSA, real-world meetings and interview scenarios usually contain multiple speakers, which are more realistic and complex. These scenarios call for a multi-speaker anonymization (MSA) system that anonymizes every speaker's voice (privacy) while keeping the anonymized voices distinctive throughout the conversation (utility).

At the time of writing, no work explores MSA due to several challenges. First, we lack evaluation metrics to assess the goodness of privacy protection and utility preservation. Second, there is no publicly available MSA tool for anonymizing conversations directly, and current speaker anonymizers used in SSA are insufficient to maintain distinctive relationships within conversations. Thus, this work aims to establish a benchmark for MSA, covering the task definition, evaluation metrics, and baseline solutions. We further discuss privacy leakage when speech from different speakers overlaps in a conversation.

The contributions of this work are as follows:
\begin{itemize}

\item We define the criteria for MSA and introduce metrics to assess its effectiveness in terms of privacy and utility, including content, naturalness, and speaker distinctiveness preservation.

\item We develop a cascaded MSA system that can handle conversations involving multiple speakers. To achieve this, we use a speaker diarization technique to aggregate the speech of each speaker and apply the spectral-clustering-based method to guarantee the correctness of speaker diarization. Following segmentation, each speaker segment undergoes individual anonymization using the disentanglement-based anonymization method with a selection-based speaker anonymizer.

\item We improve the selection-based speaker anonymizer by proposing two \textit{conversation-level} selection strategies to generate anonymized speaker vectors. Specifically, the first strategy aims to preserve the relationships between different pairs of speakers in the anonymized conversation as closely as possible to those in the corresponding original conversation, while the second aims to reduce the overall similarity among anonymized speakers. These strategies strive for the unlinkability between the original and corresponding pseudo-speaker identities for each speaker while preserving distinguishability among pseudo-speakers within a conversation.

\item We validate the effectiveness of the proposed MSA systems on both simulated and real-world non-overlapping conversations involving various numbers of speakers and background noises. Additionally, we analyze overlapping conversations for potential privacy leakage and propose possible lightweight solutions.

\end{itemize}

\begin{table}[t]
\centering
\small
\vspace{-2mm}
\caption{ Comparison of single- and multi-speaker anonymization.}
\label{tab:diff}
\begin{tabular}{m{1.2mm} m{1.5cm} | m{2.3cm} m{3cm}}
\toprule

\multicolumn{2}{c|}{Goals} & \multicolumn{1}{c}{\textbf{SSA}} & \multicolumn{1}{c}{\textbf{MSA}} \\ \midrule
\multicolumn{2}{c|}{Input} & \cellcolor{gray!10} Single-speaker original speech & \cellcolor{gray!10} Multi-speaker original conversation \\ \midrule

\multicolumn{2}{c|}{Output} & \cellcolor{gray!10} Single-speaker anonymized speech & \cellcolor{gray!10} Multi-speaker anonymized conversation \\ \midrule
\multicolumn{2}{c|}{Privacy} & \multicolumn{2}{c}{Conceal each original speaker’s identity} \\ \midrule

\multirow{3}{*}[-1em]{\rotatebox{90}{Utility}} 
& Content & \multicolumn{2}{c}{Maintain content} \\ \cmidrule(lr){3-4}
& Naturalness & \multicolumn{2}{c}{Maintain naturalness} \\ \cmidrule(lr){3-4}
& Speaker distinctiveness & \cellcolor{gray!10} Depends on downstream tasks & \cellcolor{gray!10} Speakers within one conversation should be distinctive, and turn-taking structure should remain consistent\\

\bottomrule

\end{tabular}
\vspace{-2mm}

\end{table}

\section{Related Work on \\ 
Single-speaker Anonymization}
This section reviews SSA, which serves as the foundation for this study. It describes the goals outlined in VPC~\cite{tomashenko2021voiceprivacy, tomashenko2022voiceprivacy, tomashenko2024voiceprivacy} as well as traditional and advanced SSA approaches. 

\subsection{Goals of Single-speaker Anonymization}
\label{sec:ssa-criteria}
The goals for SSA are listed on the left side of Table \ref{tab:diff}. The input of SSA is single-speaker original speech, and the output is anonymized speech. Specifically, VPC treats the SSA task as a game between users and attackers. Suppose users publish their anonymized speech after applying the SSA system to their original private speech, which involves only one speaker. This anonymized speech should conceal speaker identity when facing different attackers while keeping other characteristics unchanged to maintain intelligibility and naturalness, enabling downstream tasks to be achieved.

For privacy evaluation, assume attackers have different levels of prior knowledge about the speaker anonymization approach applied by the users. The attackers then use this prior knowledge to determine the speaker identity in the users' anonymized speech.

For utility evaluation, the primary downstream task for anonymized speech was automatic speech recognition (ASR) model training, where preserving speech content, intelligibility, and naturalness in anonymized speech is paramount. The other utility metrics, such as speaker distinctiveness preservation, depend on downstream tasks. For example, when using SSA to generate a privacy-friendly synthetic automatic speaker verification (ASV) dataset \cite{miao2024synvox2}, speaker distinctiveness should be preserved. This means that the anonymized voices of all speakers must be distinguishable from each other and should not change over time. Hence, \textit{speaker-level anonymization} \cite{tomashenko2021voiceprivacy} is applied, where all utterances from the same speaker in the dataset are converted to the same pseudo-speaker, while utterances from different speakers have different pseudo-speakers. This process requires the original speaker labels.

In VPC 2024 \cite{tomashenko2024voiceprivacy}, in addition to ASR, emotion analysis is considered as another downstream task. Preserving emotion traits in anonymized speech became essential while speaker distinctiveness is not necessary. In line with the considered application scenarios, \textit{utterance-level anonymization} \cite{tomashenko2024voiceprivacy} is applied, where each utterance is assigned to a pseudo-speaker independently of other utterances. The pseudo-speaker assignment process does not rely on speaker labels, typically resulting in a different pseudo-speaker for each utterance. Note that assigning a single pseudo-speaker to all utterances also satisfies this definition.

\subsection{Single-speaker Anonymization Approaches}
\subsubsection{Single-speaker anonymization approaches from VPC}
SSA approaches can be categorized into digital signal processing (DSP)-based and disentanglement-based methods. A DSP-based method, for example the VPC baseline \cite{patino21_interspeech}, conceals the perceived original speaker's identity by warping the spectral envelope via McAdams coefficients. However, those methods distort the content more severely than disentanglement-based methods \cite{tomashenko2021voiceprivacy,tomashenko2022voiceprivacy,tomashenko2024voiceprivacy}.

Two hypotheses underlie the disentanglement-based approaches. First, speech can be decomposed into content, speaker identity, and prosodic representations, where speaker identity is time-invariant across the utterance, while others are time-variant. 
Second, the speaker identity representation carries most of the speaker's private information. 
By modifying the original speaker representation and using it and the original content and prosodic representations to reconstruct the anonymized speech, we expect to hide most of the speaker identity information in the original speech.

Because disentanglement-based SSAs show superior effectiveness in preserving speaker privacy and maintaining utility, the majority of speaker anonymization studies have adopted a similar disentanglement-based framework, with improvements made from several perspectives\footnote{We summarize improvements into different subcategories, highlighting recently proposed novel and impactful methods. Some papers are cited multiple times for their contributions from various perspectives.}.

\subsubsection{Improved speech disentanglement}
Several works argue that disentangled content and prosodic representations still contain speaker information. To suppress leaked speaker information, various approaches have been proposed, for instance, adding differentially private noise to pitch and linguistic content \cite{shamsabadi2022differentially}, modifying pitch and speech duration \cite{mawalim2022speaker}, and applying vector quantization to content representation \cite{champion2022disentangled, champion2023anonymizing}. Recently, a neural audio codec-based approach was proposed \cite{10447871}, effectively bottlenecking speaker-related information to enhance privacy protection. \cite{yao2024musa} employed a serial disentanglement approach to better separate global speaker identity representation, linguistic content, and prosody.

\subsubsection{Improved speaker anonymizer}
A widely-used selection-based speaker anonymizer \cite{Srivastava2020DesignCF,srivastava2022privacy} replaced an original speaker vector with a mean vector (pseudo-speaker vector), which is an average of speaker vectors of randomly selected from an external pool of English speakers. Previous research \cite{tomashenko2021voiceprivacy,tomashenko2022voiceprivacy} has demonstrated that anonymized voices generated by selection-based anonymizers have limited variability due to the average operation. 
Facing the limitation, one of the top SSA submissions to VPC 2022 \cite{meyer2023prosody} utilized a generative adversarial network (GAN) to generate the anonymized speaker vectors from random noise.
\cite{miao2023language} uses an orthogonal householder neural network (OHNN) to rotate the original speaker vectors into anonymized ones, while ensuring the anonymized vectors follow the overall distribution of the original ones but do not overlap with them. 
Another work \cite{yao2024distinctive} applies singular value decomposition over a matrix composed from the speaker vectors. All these methods replace the selection-based anonymizer using linear or non-linear generative models.

\subsubsection{Flexible attribute anonymization}
SSA, as defined by VPC, focuses solely on removing speaker identity from the original speech. The linguistic content and other paralinguistic attributes remain unchanged, even though many of them, for example, age, gender, emotion, and dialect,
could potentially disclose a speaker's privacy, including geographical background, social identity, and health status \cite{zhang2023voicepm,noe2023hiding}. Researchers have explored techniques aimed at protecting voice privacy by concealing various privacy-related characteristics, such as age, gender, and accent from speech signals.

\subsubsection{Language-robust speaker anonymization}

Another area of research focuses on developing an SSA solution that can be applied to speech in unseen languages. Self-supervised learning (SSL)-based speaker anonymization has been proposed \cite{miao22_odyssey, miao2023language, zhang2023voicepm,yao2024musa}, utilizing an SSL-based content encoder to extract general context representations regardless of the input speech's language. The entire system requires no text labels or other language-specific resources, enabling it to anonymize speech data from unseen languages.

Despite advancements in SSA techniques, if we apply them directly to speech conversation data, they transform multi-speaker interactions into a single pseudo-speaker, erasing crucial turn-taking information essential for MSA. To preserve conversational dynamics while anonymizing individual speakers in MSA, one possible approach is to leverage SSA techniques as a foundation, adapting and enhancing them through additional modules to create a tailored solution. In this work, we select SSL-based SSA as the backbone, as it has been verified to maintain good intelligibility, naturalness, and applicability across multiple languages \cite{miao22_odyssey,miao2022analyzing,miao2023language}. This will be detailed in Section \ref{sec:spk-anon}.

\begin{figure*}[ht]
\centering
\includegraphics[width=6in]{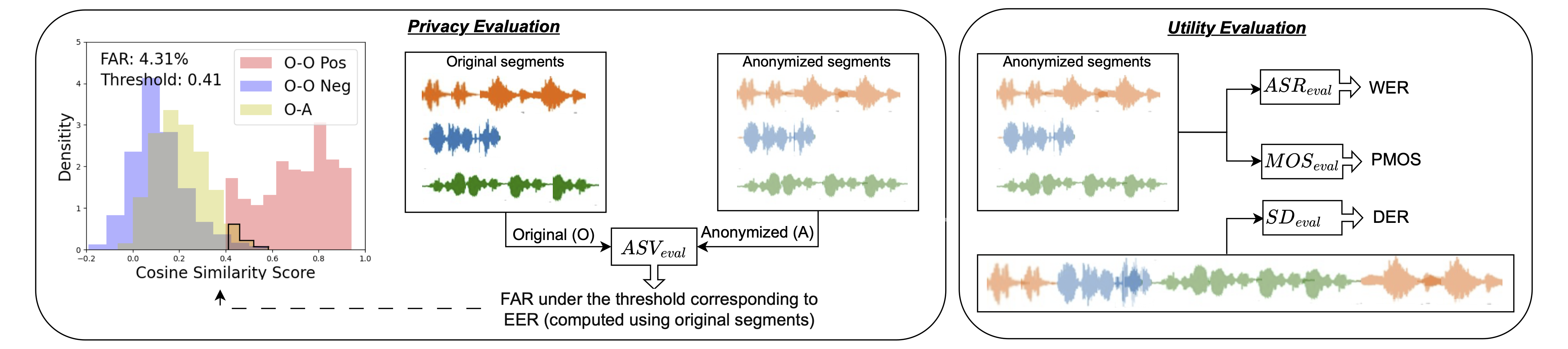}

\caption{Privacy and utility evaluation for MSA. FAR metric assesses privacy, and WER, PMOS, and DER metrics assess utility. For FAR computation, "O-O pos" represents both the enrollment and test segments being from the same original speaker, while "O-O neg" represents those from different original speakers. "O-A" represents the enrollment segment being from the original speaker and the test segment being the corresponding anonymized segment. FAR is the ratio of the black lines outlined area to the yellow area.}
\vspace{-2mm}
\label{fig:eval}

\end{figure*}

\section{Goals and Evaluation Metrics for Multi-speaker Anonymization}
This section defines the goals of MSA, along with highlighting the differences from the SSA goals described in Section \ref{sec:ssa-criteria}. Various metrics are then established to assess MSA effectiveness in terms of privacy and utility.

\subsection{Comparison of Goals for Single-speaker and Multi-speaker Anonymization}
\label{sec:requirment}
An ideal MSA system should ensure safeguarding each speaker's privacy and preserving content and naturalness, which is similar to SSA. Additionally, it should maintain the original number of speakers and the conversational turn-taking structure, thereby accurately conveying the context. Table \ref{tab:diff} summarizes and compares the goals for SSA and MSA across four perspectives: input type, output type, privacy, and utility.

The common objectives includes privacy protection, speech content preservation, and speech naturalness preservation. The differences are highlighted in grey cells. For input type, SSA takes a single-speaker original speech, while MSA takes a multi-speaker original conversation. Accordingly, their output types are a single-speaker anonymized speech and a multi-speaker anonymized conversation, respectively. In terms of speaker distinctiveness, SSA highlights that the preservation of speaker distinctiveness depends on downstream tasks involving a single speaker in each utterance, such as ASV or emotion recognition as described in Section \ref{sec:ssa-criteria}. However, MSA emphasizes preserving speaker distinctiveness within a single conversation, especially in scenarios where speaker labels are unknown, ensuring that all segments from the same speaker are attributed consistently to a single pseudo-speaker while maintaining the original number of speakers before and after anonymization to retain the logical context of the conversation.

As the input types are different, unfortunately, both \textit{utterance-level} and \textit{speaker-level anonymization} used in SSA cannot be directly applied to MSA. For example, an input conversation with multiple speakers' voices would be anonymized into a single pseudo-voice, resulting in a complete loss of turn-taking information from the original conversation. Therefore, it is essential to identify segments from different speakers and anonymize each speaker's segments individually to keep the utility of the anonymized conversation. The proposed solutions will be illustrated in Sections \ref{sec:spk-anon} and \ref{sec:spk-method}.

\subsection{Evaluation Metrics}
\label{sec:eval}
The Fig. \ref{fig:eval} illustrates the evaluation process for MSA. We use one privacy metric to assess privacy protection and three utility metrics to assess content preservation, naturalness, and speaker distinctiveness in anonymized conversations, respectively. All metric computations rely on pre-trained evaluation models. Notably, except for speaker distinctiveness, which is assessed on anonymized conversations, the computation of other metrics is conducted on single-speaker segments aggregated using diarization results from both the original and anonymized conversations.

\subsubsection{Privacy metrics}
To evaluate the effectiveness of speaker privacy protection in anonymized speech, we take the viewpoint of the attacker who uses an ASV model ($ASV_\text{eval}$) to guess the speaker identity from the anonymized speech. Given an unanonymized (original) reference utterance from a targeted user, the attacker uses the ASV model to measure how similar an anonymized utterance is to the reference in terms of speaker identity. The similarity (i.e., the ASV score) should be low for a well anonymized utterance. However, if there is some leakage of the speaker identity in the anonymized utterance, the ASV score may be higher than the ASV threshold, leading the attacker to accept the hypothesis that the anonymized utterance and the reference are uttered from the same speaker. The `success' rate of the attacker's guess is equivalent to the ASV false accept rate (FAR). The pair of an anonymized utterance and unanonymized reference is considered to be \emph{negative} data, while that of an unanonymized utterance and reference from the same speaker is considered to be \emph{positive} data. Similar metrics have been used in other security fields, e.g., membership inference attack \cite{carliniMembership2022,nasr2023tight}.

To further explain the FAR, we first define three types of enrollment and test pairs that are used to compute the FAR. Given $M$ original (O) and anonymized (A) conversations, $C_o^m$ is the $m$-th original conversation with ${N_m}$ speakers and $m \in [1, M]$. $\mathbf{x}_{\text{o}}^{mn}$ is the aggregated single-speaker segment for speaker $n$ in original conversation $C_o^m$. Similarly, $C_a^m$ is the $m$-th anonymized conversation. $\mathbf{x}_{\text{a}}^{mn}$ is the aggregated single-speaker segment for speaker $n$ in anonymized conversation $C_a^m$.

\begin{itemize}
\item {O-O positive pairs}: $\mathbf{x}_{\text{o}}^{mn}$ is split in half, denoted by $\mathbf{x}_{\text{o}}^{mn(1)}$ and $\mathbf{x}_{\text{o}}^{mn(2)}$, to form positive pairs for each conversation, and then traverses all conversations. This traversal process uses the union of sets $\bigcup$ to encompass all positive pairs:
 \begin{equation}
 P_{\text{positive}} = \bigcup_{m=1}^{M} \bigcup_{n=1}^{N_m} \{ (\mathbf{x}_{\text{o}}^{mn(1)}, \mathbf{x}_{\text{o}}^{mn(2)}) \},
 \end{equation}
\item {O-O negative pairs}: different speaker segments from each conversation that traverse all conversations form negative pairs :
 \begin{equation}
 P_{\text{negative}} = \bigcup_{m=1}^{M} \bigcup_{\substack{n=1 \\ k \neq n}}^{N_m} \{ (\mathbf{x}_{\text{o}}^{mn}, \mathbf{x}_{\text{o}}^{mk}) \}.
 \end{equation}

\item {O-A pairs}: the original single-speaker segment and its corresponding anonymized segment form as O-A pairs:
 \begin{equation}
 P_{\text{O-A}} = \bigcup_{m=1}^{M} \bigcup_{n=1}^{N_m} \{ (\mathbf{x}_{\text{o}}^{mn}, \mathbf{x}_{\text{a}}^{mn}) \}.
 \end{equation}
\end{itemize}

$ASV_\text{eval}$ is then utilized to select a threshold corresponding to the equal error rate (EER), where the FAR and the false rejection rate (FRR) are equivalent, using $P_{\text{positive}}$ and $P_{\text{negative}}$ from the original conversation. Subsequently, $P_{\text{O-A}}$ is input to $ASV_\text{eval}$ to calculate cosine similarity and compute the FAR. This FAR is determined as the number of false acceptances (under the threshold identified using original segments) divided by the total number of $P_{\text{O-A}}$.

A lower FAR indicates that $ASV_\text{eval}$ identifies anonymized test segments as dissimilar to their corresponding original enrollment segments, suggesting that most anonymized conversations conceal speaker identities, thereby safeguarding the speakers' privacy.

\subsubsection{Utility metrics}
\paragraph{Word error rate} To assess how well \textit{speech content} is preserved in anonymized speech, the word error rate (WER) is computed by using an ASR evaluation model denoted as $ASR_\text{eval}$. A lower WER, similar to that of the original speech, indicates a good speech content preservation ability.

\paragraph{Predicted mean opinion score} To assess how well \textit{speech naturalness} is preserved in anonymized speech, the predicted mean opinion score (PMOS) is computed by a mean opinion score (MOS) prediction network \cite{cooper2021generalization} denoted as $MOS_\text{eval}$. A higher PMOS, similar to that of the original speech, indicates a good speech naturalness preservation ability.

\paragraph{Diarization error rate} To assess how well the \textit{speaker distinctiveness} is preserved in anonymized conversations, the diarization error rate (DER) is computed by a speaker diarization (SD) evaluation model denoted as $SD_\text{eval}$. Anonymized conversations with similar speaking turns and speaker distinctiveness to the original speech will achieve a DER similar to the original ones. Conversely, higher or lower DER than those of original conversations means worse or better distinctiveness preservation, respectively.

\begin{figure}[!t]
\centering
{\includegraphics[width=3.5in]{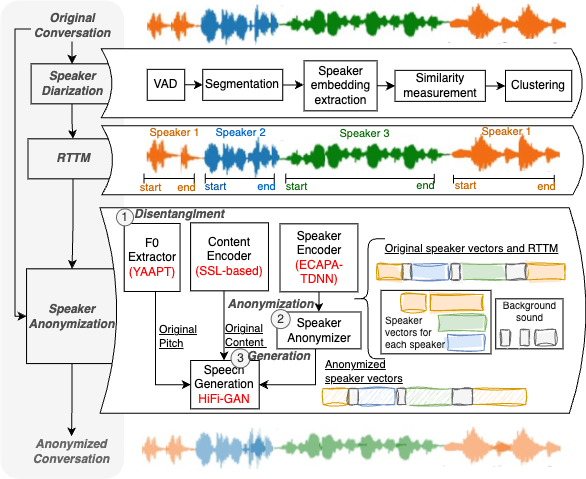}}
\caption{Pipeline of cascaded MSA, where the SD module is first used to aggregate single-speaker segments, followed by disentanglement-based anonymization for individual anonymization. }
\vspace{-2mm}
\label{fig:main_structure}
\end{figure}

\section{A Cascaded Multi-speaker \\ Anonymization System}
Given the absence of publicly accessible MSA tools for users to directly use in anonymizing conversations, this section addresses this gap by proposing a cascaded MSA system, as shown in Fig. \ref{fig:main_structure}.

The original conversation speech is fed into the SD module to generate rich transcription time-marked (RTTM) information for each speaker. The RTTM data is crucial as it serves as a foundation for aggregating the speech of each speaker and reconstructing the conversation sequentially. Subsequently, the individual single-speaker speech aggregated on the basis of RTTM data is anonymized to the same pseudo-speaker, while the speech from different speakers is anonymized into distinct pseudo-speakers\footnote{\textit{speaker-level anonymization}}. Notably, the background audio remains unaltered to preserve the authenticity and realism of the conversation. In this work, the widely-used spectral-based SD and the SSL-based SSA approach with selection-based anonymizer are chosen to establish a basic MSA framework\footnote{Note that this pipeline is not limited to a specific SD and speaker anonymization method.}. The details of each component are provided in the following.

\subsection{Speaker Diarization}
The spectral clustering-based SD system involves multiple stages \cite{shum2013unsupervised,sell2014speaker, snyder2019speaker}, as illustrated in Fig. \ref{fig:main_structure}. First, a voice activity detection (VAD) system is used to filter out the non-speech regions. The active speech regions are then split into short fixed-length segments with a specific overlapping ratio. Subsequently, a pre-trained speaker embedding extractor is used to extract speaker vectors for each segment. Following this, a scoring backend, such as cosine scoring or probabilistic linear discriminant analysis (PLDA) \cite{rajan2014single,villalba2013handling}, is applied to compute similarity scores between pairs of segments. A clustering algorithm \cite{sell2018diarization,wang2018speaker,lin2019lstm} is then used to assign a unique speaker label to each segment. Finally, the clustering results are summarized into the RTTM file, which contains the start time, duration, and speaker ID for each talking segment, providing the foundation for separating the speech of individual speakers.

\begin{figure}[!t]
\centering
{\includegraphics[width=3.5in]{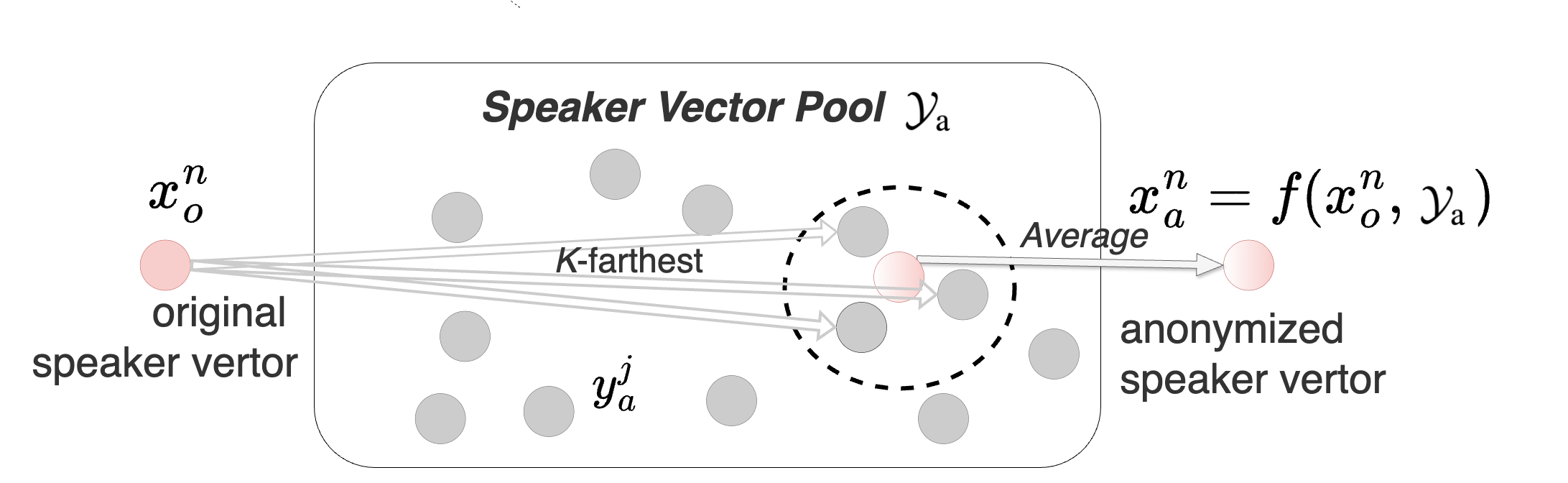}}
\caption{Workflow of selection-based speaker anonymizer using an external speaker vector pool, adopted by VPC baseline systems. The input speaker vector $\boldsymbol{x}_{o}^{n}$ is anonymized by selecting $K$-farthest vectors in the pool $\symbolset{Y}_\text{a} = \{\symbolvec{y}_\text{a}^1, \ldots, \symbolvec{y}_\text{a}^P\}$. The anonymized output $\symbolvec{x}_{a}^n$ is set to be the average of the $K$-farthest vectors. }
\vspace{-2mm}

\label{fig:spk-anon}
\end{figure}

\subsection{Speaker Anonymization}
\label{sec:spk-anon}
Considering the goals for achieving MSA, we select the SSL-based SSA as the backbone as it has been verified to maintain good intelligibility and naturalness, and can be used for multiple languages \cite{miao22_odyssey,miao2022analyzing,miao2023language}. After using the RTTM to aggregate the speech of each speaker and background audio, the speaker anonymization system anonymizes the speech of each speaker separately. Specifically, the system involves three steps, as shown at the bottom of Fig. \ref{fig:main_structure}.

\noindent
\textit{Original speech disentanglement}: The first step aims to disentangle the original speech into different components representing various speech attributes. This includes extracting frame-level content features via SSL-based content encoders \cite{miao22_odyssey, miao2022analyzing}, frame-level F0 using the YAAPT algorithm \cite{kasi2002yet}, and segment-level original speaker vectors individually for each speaker via ECAPA-TDNN \cite{desplanques2020ecapa} models. This helps separate speaker identity information from linguistic and prosodic content, facilitating the concealment of speaker identity in the following step. This step aims to modify the speaker vectors for anonymization.

\noindent
\textit{Speaker anonymizer}: Hiding the original speaker's identity for each speaker is crucial. Most works focus on modifying the original speaker embeddings using a speaker anonymizer, assuming that identity is mainly encoded in them. 
Given one conversation with $N$ speakers, let us denote their original speaker vectors as $ \symbolset{X}_\text{o}= \{\symbolvec{x}_\text{o}^1, \ldots, \symbolvec{x}_\text{o}^N\}$, where $\symbolvec{x}_\text{o}^n \in \mathbb{R}^D$ is the $D$-dimensional segment-level speaker vector of the $n$-th speaker\footnote{Each speaker has multiple segments in a conversation. These segments are aggregated into one single-speaker speech on the basis of diarization results, and then the speaker vector for each speaker is extracted, denoted $\symbolvec{x}_\text{o}^n$.}.
Let us then define an external pool of speaker vectors from $P$ speakers as  $\symbolset{Y}_\text{a} = \{\symbolvec{y}_\text{a}^1, \ldots, \symbolvec{y}_\text{a}^P\}$, where $\symbolvec{y}_\text{a}^p\in\mathbb{R}^{D}, \forall p\in[1, P]$. 
Each anonymized speaker vector $\symbolvec{x}_\text{a}^n = f(\symbolvec{x}_\text{o}^n, \symbolset{Y}_\text{a})$ is obtained for speaker $n \in [1, N]$ by the speaker anonymizer $f$.
One commonly-used speaker anonymizer in VPC relies on an external pool as shown in Fig. \ref{fig:spk-anon}.
The anonymization algorithm in VPC defines a function $f_\text{VPC}: \mathbb{R}^{D}\times\mathbb{R}^{D\times{P}} \rightarrow\mathbb{R}^{D}$, which produces an anonymized vector $\symbolvec{x}_\text{a}^n = f(\symbolvec{x}_\text{o}^n, \symbolset{Y}_a), \forall{n}\in[1, N]$.  
Note that the external pool $\symbolset{Y}_\text{a}$ is sourced from speakers different from the $N$ speakers to be anonymized and $P \gg N$.\footnote{To protect the identity of the speakers in the external pool, we average the ten most similar, gender-consistent speaker vectors along with the original pool speaker vector itself to generate a replacement for this pool speaker vector. This differs from the usual settings where the speaker vectors in the external pool are left unanonymized.} The anonymizer function $f$ in the VPC processes each original speaker vector $\symbolvec{x}_\text{o}^n$ individually by searching for the $K$ farthest, same-gender speaker vectors in the external speaker vector pool and averaging them to generate the anonymized speaker vector $\symbolvec{x}_\text{a}^n$ sequentially.
\noindent
\textit{Anonymized speech generation}: The anonymized speaker vector, original prosody, and content features are then fed into the neural vocoder HiFi-GAN \cite{kong2020hifi} to generate individual anonymized segments. These segments are reconstructed into an anonymized conversation utilizing the temporal information provided in the RTTM file.

\begin{figure}[!t]
\centering
\includegraphics[width=3.7in]{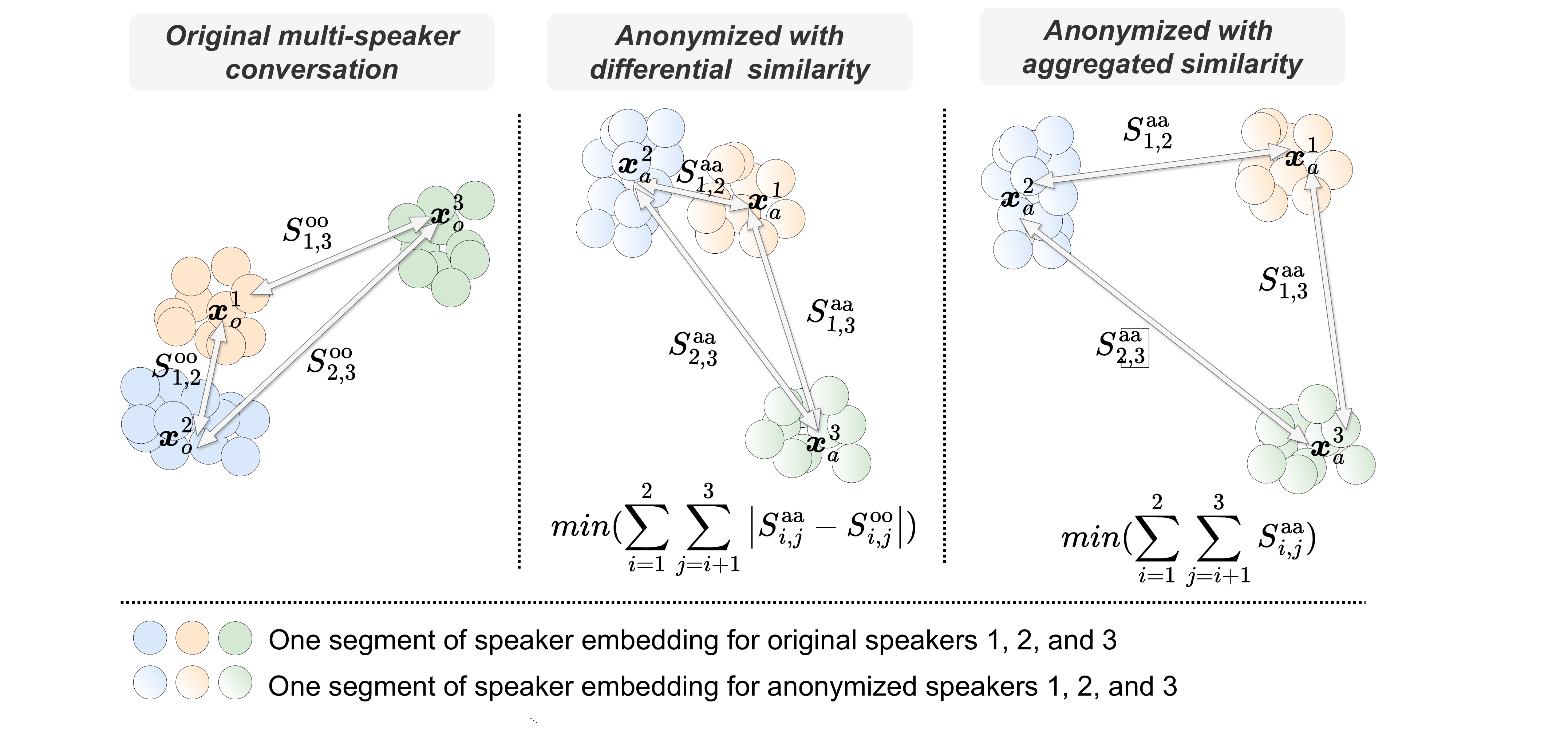}
\caption{Illustration of proposed differential and aggregated similarity-based anonymized speaker vector selection methods for $N=3$ speakers. Differential similarity constraints (middle) maintain original relationships (left), while aggregated similarity constraints (right) maximize speaker differentiation.}
\label{fig:as-ds}
\vspace{-2mm}

\end{figure}

\section{Conversation-level Speaker Anonymizer for Multi-speaker Anonymization}
\label{sec:spk-method}
The cascaded MSA described in Section \ref{sec:spk-anon}, utilizing existing modules, i.e., SD and SSA, can anonymize conversations. However, the speaker anonymizer designed for SSA only considers speaker privacy protection, lacking consideration for speaker distinctiveness in conversations. This section will enhance the speaker anonymizer of the cascaded MSA by proposing two \textit{conversation-level anonymization} approaches.

\begin{algorithm}[!ht]
\footnotesize
\caption{Anonymization function $f_\text{MSA}$}
\label{alg:proposed}

\KwData{$ \symbolset{X}_\text{o}= \{\symbolvec{x}_\text{o}^1, \ldots, \symbolvec{x}_\text{o}^N\}$: original speaker vectors \\
 \hspace {20pt} 
 $\symbolset{Y}_\text{a} = \{\symbolvec{y}_\text{a}^1, \ldots, \symbolvec{y}_\text{a}^P\}$: external pool
 \\
 \hspace{20pt} $L_{\text{far}}$: number of farthest speaker vectors \\
 \hspace {20pt} $L_{\text{prune}}$: number of speaker vector choices to keep during greedy search \\
}
\tcp{Similarity matrices}
$\symbolmat{S}^{x_oy_a} \leftarrow$ compute similarity matrix given $ \symbolset{X}_\text{o}$ and $ \symbolset{Y}_\text{a}$. \\
$\symbolmat{S}^{x_ox_o} \leftarrow$ compute similarity matrix given $ \symbolset{X}_\text{o}$. \\ $\symbolmat{S}^{y_ay_a} \leftarrow$ compute similarity matrix given $ \symbolset{Y}_\text{a}$. \\
\tcp{Matrix to save the indices of farthest speakers}
$\symbolmat{D}\in \mathbb{R}^{N\times L_{\text{far}} }$ \\

\tcp{Protecting privacy}
\ForEach{$i \in [1, N]$} {
 \tcp{for each speaker, choose $L_{\text{far}}$ farthest speaker vectors from pool as candidate speaker vectors}
 $\symbolmat{D}[i]$ = $\arg\text{sort}(\symbolmat{S}^{x_oy_a}[i])[:L_{\text{far}}]$

}

\tcp{Maintaining utility}
\tcp{Buffer to save external candidate speaker index and similarity}
$\symbolvec{s} = \Big[([D_{1,1}],0.0), ([D_{1,2}],0.0), \cdots, ([D_{1,L_\text{far}}], 0.0)\Big]$\\
$\tilde{\symbolvec{s}} = []$ \\
\ForEach{$i \in [2, N]$} {
 \ForEach{$(\symbolvec{l}, s)$ in $\symbolvec{s}$} {
 \tcp{$\symbolvec{l}$: 
 a list of speaker indices and represents one possible speaker vector choice for previous i-1 speakers}
 \tcp{$s$: the similarity score}
 \ForEach{$j$ in $\symbolmat{D}[i]$} {
 \Case{Use $\mathcal{L}$ in Eq.(\ref{eq:l_as})}{
 \tcp{ speaker distinctiveness preservation: the sum of cosine similarities is minimum}
 \ForEach{$k$ in $\symbolvec{l}$}{
 $s = s + S_{j,k}^{x_ax_a}$ 
 }
 }
 \Case{Use $\mathcal{L}$ in Eq.(\ref{eq:l_ds})}{
 \tcp{speaker distinctiveness preservation: the sum of cosine similarities between X and Y is minimum}
 \ForEach{$k$ in $\symbolvec{l}$}{
 $s = s + \left| S_{j,k}^{x_ax_a} - S_{j,k}^{x_ox_o} \right| $ 
 }
 }
 $\boldsymbol{l} = \boldsymbol{l} + \{j\}$ \\
 $\tilde{\symbolvec{s}} = \tilde{\symbolvec{s}} + \{\boldsymbol{l}, s\}$\\ 
 }
 }
 \tcp{sorting based on the value of $s$}
 $\tilde{\symbolvec{s}} = \text{sort}(\tilde{\symbolvec{s}})$ \\
 \tcp{update the statistics and keep the $L_\text{prune}$ choices with the smallest similarities}
 $\symbolvec{s} = \tilde{\symbolvec{s}}[:L_\text{prune}]$
}

\ForEach{$i$ in $\symbolvec{s}[N][0][0]$}{
\tcp{the list $\symbolvec{s}$ has length N, each element $\symbolvec{s}[i]$ has length $L_\text{prune}$, the first element $\symbolvec{s}[i][0]$ is a tuple $(\symbolvec{l}, s)$, where $\symbolvec{l}$ is the list of $i$ speakers indices under the minimum similarity s. Likewise, $\symbolvec{s}[N][0][0]$ is N speaker indexes under the minimum similarity.}

 \tcp{Retrieve the index for each speaker}
 $k = \symbolvec{s}[N][0][0][i]$ \\
 \tcp{Assign the selected external vector}
 $\symbolvec{x}_\text{a}^{i} = \symbolvec{y}_\text{a}^{k}$
}
 \KwOut {Anonymized vectors $\symbolset{X}_\text{a}=\{\symbolvec{x}_\text{a}^{1}, \cdots, \symbolvec{x}_\text{a}^{N}\}$ }

\end{algorithm}

Different from SSA where the anonymizer $f_\text{VPC}$ process each $\symbolvec{x}_o^{n}$ independently, we aim to design a more complicated $f_{\text{MSA}}: \mathbb{R}^{D\times{N}}\times\mathbb{R}^{D\times{P}} \rightarrow\mathbb{R}^{D\times{N}}$ so that the anonymized speaker vectors well conceal the speakers' identities and stay distinctive across the $N$ speakers. 

Before explaining how the goodness of anonymization is measured, let us define a similarity matrix $\symbolmat{S}^{x_oy_a}\in\mathbb{R}^{N\times{P}}$, where the element ${S}^{x_oy_a}_{i,j}$ is equal to the similarity between the original speaker vector $\symbolvec{x}_\text{o}^i$ and the candidate anonymized vector $\symbolvec{y}_\text{a}^j$. A common choice is to compute the cosine similarity, i.e.,
\begin{align}
{S}^{x_oy_a}_{i,j} = \frac{{\symbolvec{x}_\text{o}^i}^\top\symbolvec{y}_\text{a}^j}{
\sqrt{{\symbolvec{x}_\text{o}^i}^\top\symbolvec{x}_\text{o}^i \cdot {\symbolvec{y}_\text{a}^j}^\top\symbolvec{y}_\text{a}^j}}.
\end{align}
Similarly, we define $\symbolmat{S}^{y_ay_a}\in\mathbb{R}^{P\times{P}}$ that measure the similarities among candidate anonymized speaker vectors.

Suppose that the anonymized speaker vector {$\symbolvec{x}_a^n$ has been obtained for each speaker $n\in[1,N]$ by the multi-speaker anonymizer $\symbolset{X}_a=f_{\text{MSA}}(\symbolset{X}_o, \symbolset{Y}_a)$}. The similarities between the original and selected anonymized speaker vectors can be represented:
\begin{align}
{S}^{x_ox_a}_{i,j} = \frac{{\symbolvec{x}_\text{o}^i}^\top\symbolvec{x}_\text{a}^j}{
\sqrt{{\symbolvec{x}_\text{o}^i}^\top\symbolvec{x}_\text{o}^i \cdot {\symbolvec{x}_\text{a}^j}^\top\symbolvec{x}_\text{a}^j}}.
\end{align}
Additionally, $\symbolmat{S}^{x_ox_o}\in\mathbb{R}^{N\times{N}}$ and $\symbolmat{S}^{x_ax_a}\in\mathbb{R}^{N\times{N}}$ are defined to measure the similarities among original and selected anonymized speaker vectors, respectively.

As explained in Section \ref{sec:requirment}, in an ideal MSA, pseudo-speakers should meet two criteria:
\begin{itemize}[]
\item \textbf{Protecting privacy}: to hide the original speaker identity, an original speaker vector and its anonymized version should be dissimilar. This means that ${S}^{x_ox_a}_{i,i}$ should be small for $\forall i \in[1, N]$ and the sum of these similarities can be represented as $\sum_{i=1}^{N} {S}^{x_ox_a}_{i,i}$.
\item \textbf{Maintaining utility}: to maintain speaker distinctiveness after anonymization, ${S}^{x_ax_a}_{i,j}$ should be small for $\forall i\neq j$. Two approaches are proposed to achieve good utility.

\begin{itemize}
\item[$\bullet$] Differential similarity (DS): This approach maintains the utility by minimizing the difference between the similarity of the original speaker pair (e.g., $\symbolvec{x}_\text{o}^i$ and $\symbolvec{x}_\text{o}^j)$ and that of the corresponding anonymized speaker pair (e.g., $\symbolvec{x}_\text{a}^i$ and $\symbolvec{x}_\text{a}^j$), calculated as $\sum_{i=1}^{N-1}\sum_{j=i+1}^{N} \left| {S}^{x_ax_a}_{i,j} - {S}^{x_ox_o}_{i,j} \right|$.
\item[$\bullet$] Aggregated similarity (AS): This approach directly minimizes the similarities across anonymized speakers ${S}^{x_ax_a}_{i,j}$, to achieve better differentiation between speakers, calculated as $\sum_{i=1}^{N-1}\sum_{j=i+1}^{N} {S}^{x_ax_a}_{i,j}$.
\end{itemize}

\end{itemize}

Fig. \ref{fig:as-ds} illustrates the aforementioned DS and AS approaches when $N=3$. Combining both privacy and utility constraints leads to two loss functions:
\begin{align}
\mathcal{L}^{\text{DS}}(\symbolset{X}_\text{o}, \symbolset{Y}_\text{a}, f_{\text{MSA}}) & = \sum_{i=1}^{N} {S}^{x_ox_a}_{i,i} + \sum_{i=1}^{N-1}\sum_{j=i+1}^{N} \left| {S}^{x_ax_a}_{i,j} - {S}^{x_ox_o}_{i,j} \right|,
\label{eq:l_ds}
\end{align}
and
\begin{align}
\mathcal{L}^{\text{AS}}(\symbolset{X}_\text{o}, \symbolset{Y}_\text{a}, f_{\text{MSA}}) & = \sum_{i=1}^{N} {S}^{x_ox_a}_{i,i} + \sum_{i=1}^{N-1}\sum_{j=i+1}^{N} {S}^{x_ax_a}_{i,j}.
\label{eq:l_as}
\end{align}

To minimize the loss function (either Eq.(\ref{eq:l_ds}) or Eq.(\ref{eq:l_as})), we design an $f_{\text{MSA}}$ and describe it in the python-like Algorithm \ref{alg:proposed}. Note that we perform gender-dependent anonymization in this implementation. This is done by separating the input $\symbolset{Y}_{\text{a}}$ and $\symbolset{X}_{\text{o}}$ into gender-dependent subsets and execute Algorithm \ref{alg:proposed} separately for female and male. This guarantees that the gender of each speaker remains the same before and after anonymization. When no gender annotation is available, a pre-trained gender recognition model predicts the gender.

\section{Evaluation}
In this section, we primarily evaluate the proposed system on non-overlapping datasets using various privacy and utility metrics described in Section \ref{sec:eval}. This includes assessments on simulated datasets with different numbers of speakers, both clean and noisy speech, as well as real-world conversations. Finally, we analyze the potential privacy leakage in overlapping segments.

\begin{table}[t]
\centering
\vspace{-2mm}
\caption{ Notations for the evaluated MSA.}
\label{tab:notations}
\begin{tabular}{cccc}
\toprule
Notation & Optimization level & {Speaker anonymizer} \\\midrule
\textbf{$A_{OHNN}$} \cite{miao2023language} & {\textit{Speaker-level}} & {OHNN} \\ 
\textbf{$A_{Select}$} \cite{miao22_odyssey} &\textit{Speaker-level} & {Selection-based} \\ 
 \textbf{$A_{AS}$} & \textit{Conversation-level} & Minimum $\mathcal{L}^{\text{AS}}$ \\ 
 \textbf{$A_{DS}$} & \textit{Conversation-level} & Minimum $\mathcal{L}^{\text{DS}}$ \\ 

\bottomrule
\end{tabular}
\vspace{-2mm}

\end{table}

\subsection{System Configurations}
\subsubsection{Multi-speaker anonymization configurations}

All the MSAs examined in this work are based on cascaded structures, utilizing SD and SSL-based speaker anonymization with various speaker anonymizers, all of which are well-pretrained models. The proposed \textit{conversation-level speaker anonymizer} is a selection procedure with specific conditions, eliminating the need for additional training.

For SD, we apply an efficient and robust spectrum clustering-based approach, which is implemented with the WeSpeaker Toolkit \footnote{\url{https://github.com/wenet-e2e/wespeaker/tree/master}} \cite{wang2023wespeaker}. It first uses Silero-VAD \footnote{\url{https://github.com/snakers4/silero-vad}} to remove silent segments, then, we follow the common setting from \cite{snyder2019speaker} and split the long audio into 1.5-second segments with a 0.75-second overlap. Each segment is fed into the speaker recognition model, which uses a context-aware masking-based structure \cite{wang23ha_interspeech} pre-trained on the VoxCeleb2 dataset \cite{chung2018voxceleb2}. After that, spectral clustering \cite{shum2013unsupervised,sell2014speaker} is performed to aggregate the segments into several speakers by analyzing similarity metrics. The diarization results are saved into the RTTM file, which summarizes the timestamps of each speaker's active speech segments.

Table \ref{tab:notations} lists the notations for the different MSA approaches that were examined. \textbf{$A_{OHNN}$} and \textbf{$A_{Select}$} are the cascaded MSAs using \textit{speaker-level} speaker anonymizers designed for SSA. \textbf{$A_{DS}$} and \textbf{$A_{AS}$} are those using the proposed \textit{conversation-level} speaker anonymizers. Specifically, all approaches share the same backbone, utilizing the YAAPT algorithm \cite{kasi2002yet} to extract the fundamental frequency (F0); the ECAPA-TDNN architecture, with 512 channels in the convolutional frame layers \cite{desplanques2020ecapa}, provides 192-dimensional speaker identity representations; the HuBERT-based soft content encoder \cite{van2021comparison} uses a convolutional neural network (CNN) encoder along with the first and sixth transformer layers from the pre-trained HuBERT base model. It downsamples a raw audio signal into a continuous 768-dimensional representation, subsequently mapped to a 200-dimensional vector using a projection layer to predict discrete speech units. These units are derived by discretizing the intermediate 768-dimensional representations through \textit{k}-means clustering\footnote{\url{https://github.com/pytorch/fairseq/tree/main/examples/textless_nlp/gslm/speech2unit}} \cite{polyak2021speech,lakhotia2021generative}. The configuration of HiFi-GAN is consistent with \cite{polyak2021speech}. Additional training procedures are detailed in \cite{miao22_odyssey}.

\textbf{$A_{OHNN}$}, the state-of-the-art speaker-level speaker anonymizer that achieves good speaker distinctiveness for single-speaker utterances, uses an OHNN-based anonymizer trained on authentic VoxCeleb2, utilizing random orthogonal Householder reflections with a random seed of 50 for parameter initialization. An additive loss function is used, combining weighted angular margin softmax and cosine similarity. Training details are available in \cite{miao2023language}.

\textbf{$A_{Select}$}, the commonly-used speaker anonymizer, uses a selection-based strategy that identifies the 200 farthest same-gender speaker vectors from an external speaker vector pool (specifically LibriTTS-train-other-500 \cite{zen2019libritts}), subsequently averaging 10 randomly-selected ones as the pseudo-speaker vector\footnote{10 vectors instead of the default setting of 100 used in VPC, as it has been demonstrated that averaging 100 (large number) vectors significantly reduces the distinctiveness of anonymized speakers \cite{tomashenko2024voiceprivacy}.}.

Both \textbf{$A_{AS}$} and \textbf{$A_{DS}$} are based on the same external speaker vector pool as \textbf{$A_{Select}$}. $L_{\text{far}}=200$, $L_{\text{prune}}=10,000$. The gender recognition model used before selecting an anonymized speaker vector from the pool is a fine-tuned version of wav2vec2-xls-r-300m\footnote{\url{https://huggingface.co/facebook/wav2vec2-xls-r-300m}} on Librispeech-clean-100. It achieves 99\% accuracy on the LibriSpeech test-clean subset.

\begin{table}[t]
 \centering
  \vspace{-2mm}

 \caption{Simulated datasets statistics (min/average/max) }
 \resizebox{\textwidth}{!}{
 \begin{tabular}{cccc}
 \toprule
 \textbf{\# Spk} & \textbf{\# Utt} & \textbf{ Duration (s)} & \textbf{Speech ratio (\%)}\\
 \midrule
 2 & 1121 & 3.04 / 14.70 / 76.28 & 64.37 / 93.21 / 100.00 \\
 3 & 821 & 4.23 / 21.18 / 101.50 & 67.35 / 94.68 / 100.00 \\
 4 & 623 & 8.27 / 29.02 / 123.01 & 82.97 / 95.41 / 100.00 \\
 5 & 500 & 12.30 / 36.32 / 100.38 & 80.48 / 95.64 / 99.89\\
 \bottomrule
 \end{tabular}}
 \label{tab:lib_speech_stats}
 \vspace{-2mm}

\end{table}

\begin{table}[t]
 \centering
  \vspace{-2mm}
 \caption{VoxConverse dataset statistics (min/average/max)}
\resizebox{\textwidth}{!}{
 \begin{tabular}{cccc}
 \toprule
 \textbf{\# Spk} & \textbf{\# Utt} & \textbf{ Duration (s)} & \textbf{Speech ratio (\%)}\\
 \midrule
 1 / 2.85 / 9 & 56 & 21.99 / 149.03 / 426.14 & 10.73 / 86.66 / 99.75 \\
 \bottomrule
 \end{tabular}}
 \label{tab:vox_speech_stats}
  \vspace{-2mm}

\end{table}

\subsection{Evaluation Setup}
\subsubsection{Evaluation datasets}
\paragraph{Simulated datasets}
We simulated four different subsets using the LibriSpeech test-clean subset, which includes 5 hours of audio from 40 speakers \cite{panayotov2015librispeech}. Detailed information about the simulation data is presented in Table \ref{tab:lib_speech_stats}. Each subset contains a fixed number of speakers, with the number varying from 2 to 5 across different subsets. The total duration for each subset is 5 hours, and there is no overlap between speakers.

In addition to the clean subsets, we also simulated another four corresponding subsets augmented with background noises from the MUSAN collection \cite{musan2015}, scaled with a randomly-selected signal-to-noise ratio (SNR) from ${5, 10, 15, 20}$ dB. Furthermore, with a 50\% probability, we randomly selected room impulse responses (RIR) \cite{ko2017study} to introduce reverberation, simulating far-field audio.

\paragraph{Real-world dataset}
VoxConverse \cite{chung2020spot} is a real-world conversational dataset derived from YouTube. We selected 2.31 hours of non-overlapping conversations, each lasting less than 7 minutes, from the development set. Table \ref{tab:vox_speech_stats} provides the statistics of the selected conversations.

\subsubsection{Evaluation models}
$ASV_\text{eval}$ is the publicly available ECAPA-TDNN model\footnote{\url{https://huggingface.co/speechbrain/spkrec-ecapa-voxceleb}}, trained on VoxCeleb1 \cite{nagrani2017voxceleb} and 2 \cite{chung2018voxceleb2}. $ASR_\text{eval}$ is a model fine-tuned on LibriSpeech-train-960 \cite{panayotov2015librispeech} from {wav2vec2-large-960h-lv60-self}\footnote{\url{https://huggingface.co/facebook/wav2vec2-large-960h-lv60-self}}, using a SpeechBrain \cite{speechbrain} recipe\footnote{\url{https://huggingface.co/speechbrain/asr-wav2vec2-librispeech}}. $SD_\text{eval}$ is the same speaker diarization model used in MSA. For $MOS_\text{eval}$ \cite{cooper2021generalization}, is a model fine-tuned on the Blizzard Challenge for TTS \cite{blizzard} and the Voice Conversion Challenge \cite{das2020predictions} from {wav2vec2-base}\footnote{\url{https://huggingface.co/facebook/wav2vec2-base}} by mean-pooling the model’s output embeddings, adding a linear output layer, and training with L1 loss. We utilized the predicted MOS instead of human perception-based MOS from listening tests due to time and cost constraints. The predicted MOS is reasonably well-aligned with human perception \cite{cooper2021generalization}. In our previous work \cite{miao2023language}, we demonstrated that the ranking of the predicted MOS of original and anonymized speech, generated by different speaker anonymization systems, is consistent with those from listening tests conducted by VPC \cite{tomashenko2021voiceprivacy}. This observation holds for clean datasets, and therefore we only computed PMOS for clean simulated datasets in the following experiments.

Note that except for the DER computation, where $SD_\text{eval}$ takes the conversation as input, the other metrics' computations require RTTM to split the original and anonymized conversation into single-speaker segments. In real MSA applications, we assume there is no RTTM available. MSA produces predicted RTTM using the SD model and then uses this RTTM to split and reconstruct the audio. In our experiments, we also provide the results using real RTTM (ground truth) as the upper baseline.

\begin{figure*}[!t]
 \centering
 \subfloat[2-spk ]{\includegraphics[width=0.25\textwidth]{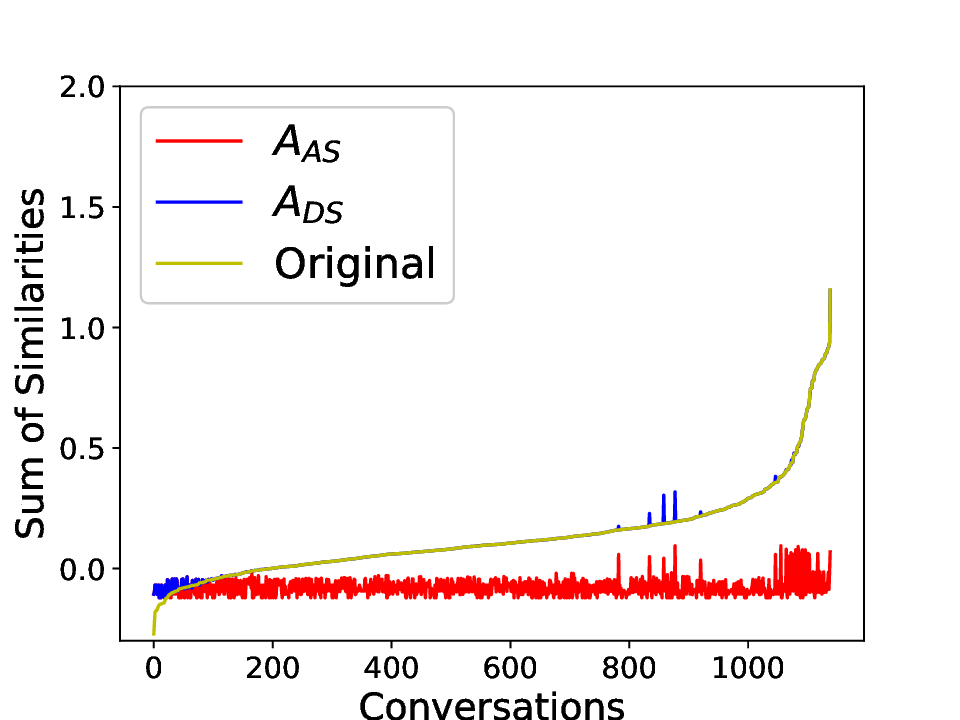}\label{fig:sub1}}
 \hfill
 \subfloat[3-spk ]{\includegraphics[width=0.25\textwidth]{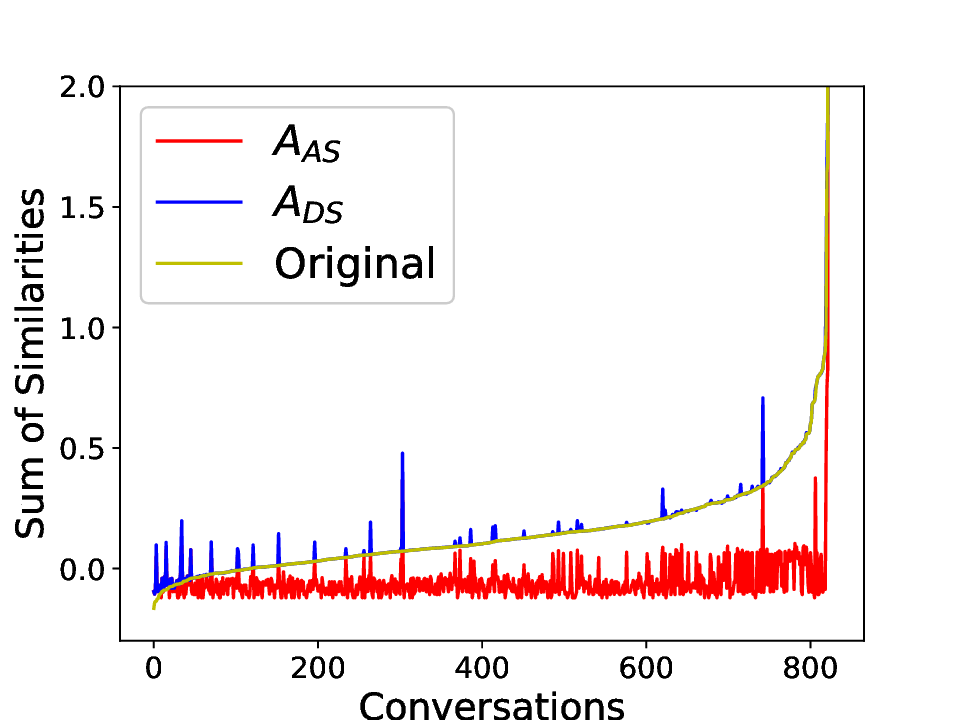}\label{fig:sub2}}
 \hfill
 \subfloat[4-spk ]{\includegraphics[width=0.25\textwidth]{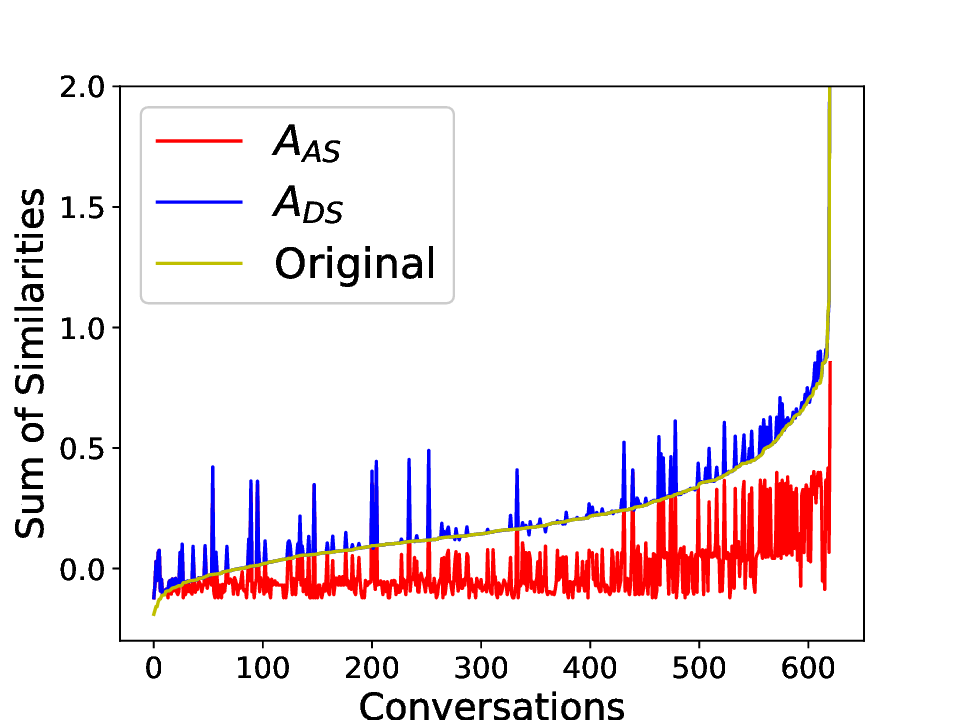}\label{fig:sub3}}
 \hfill
 \subfloat[5-spk ]{\includegraphics[width=0.25\textwidth]{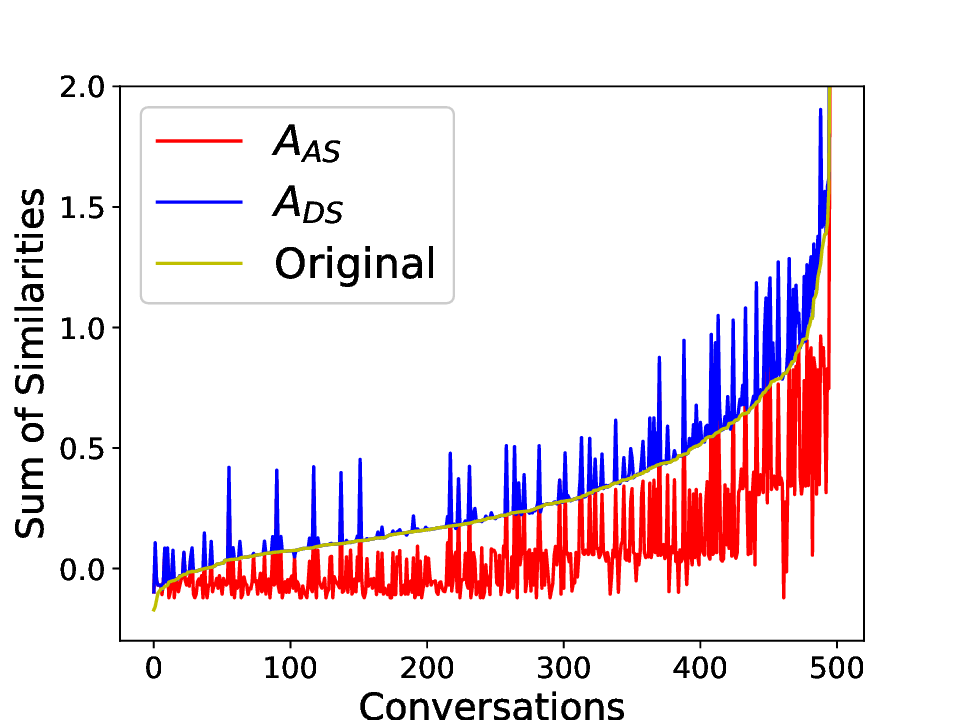}\label{fig:sub4}}
 
\caption{Sum of similarities for all combination speaker pairs per conversation, with each pair consisting of two different speakers, for the original data, $A_{AS}$, and $A_{DS}$ using predicted RTTM on clean simulation datasets.}
\vspace{-2mm}

 \label{fig:com}
\end{figure*}

\begin{table*}[]
 \centering
  \vspace{-2mm}
 \caption{FAR(\%) $ \downarrow$ on original, resynthesized, and anonymized segments on simulated conversations using the real RTTM.}
 \begin{tabular}{c|ccccc|ccccc}
 \toprule
 & \multicolumn{5}{c|}{Clean} & \multicolumn{5}{c}{Noise} \\
 & Resyn & $A_{OHNN}$ & $A_{Select}$ & $A_{DS}$ & $A_{AS}$ & Resyn & $A_{OHNN}$ & $A_{Select}$ & $A_{DS}$ & $A_{AS}$ \\
 \midrule
 2-spk & 99.52 & 2.90 & 0.04 & 3.12 & 3.12 & 96.18 & 1.98 & 0.57 & 1.36 & 1.93 \\
 3-spk & 99.35 & 2.71 & {0.40} & 2.46 & 2.46 & 95.15 & 2.18 & 0.93 & 1.33 & {1.74} \\
 4-spk & 99.48 & 3.06 & {0.24} & 2.54 & 2.70 & 94.77 & 2.54 & 0.72 & 1.61 & {1.53} \\
 5-spk & 99.56 & 3.19 & {0.48} & 2.54 & 1.98 & 94.31 & 2.42 & 1.01 & 1.53 & {1.29} \\

 \bottomrule
 \end{tabular}
 \label{tab:real-rttm-far}
  \vspace{-2mm}

\end{table*}

\begin{table*}[]
 \centering
  \vspace{-2mm}

 \caption{FAR(\%) $ \downarrow$ on original, resynthesized, and anonymized segments on simulated conversations using the predicted RTTM.}
 \begin{tabular}{c|ccccc|ccccc}
 \toprule
 & \multicolumn{5}{c|}{Clean} & \multicolumn{5}{c}{Noise} \\
 & Resyn & $A_{OHNN}$ & $A_{Select}$ & $A_{DS}$ & $A_{AS}$ & Resyn & $A_{OHNN}$ & $A_{Select}$ & $A_{DS}$ & $A_{AS}$ \\
 \midrule
 2-spk & 98.97 & 4.31 & 1.03 & 2.06 & 2.21 & 94.60 & 1.49 & 0.56 & 1.42 & 1.15 \\
 3-spk & 98.95 & 4.30 & {1.05} & 1.81 & 1.35 & 95.72 & 2.10 & 0.45 & 1.40 & {0.91} \\
 4-spk & 99.35 & 6.69 & {1.11} & 2.58 & 1.89 & 97.69 & 4.13 & 1.86 & 1.38 & {1.69} \\
 5-spk & 99.72 & 8.66 & {2.70} & 3.45 & 2.47 & 98.92 & 3.61 & 1.22 & 1.81 & {1.26} \\

 \bottomrule

 \end{tabular}
 \label{tab:pre-rttm-far}
  \vspace{-2mm}

\end{table*}

\subsection{Experimental Results on non-overlapping conversations}
\subsubsection{Examination of the proposed conversational-level speaker anonymizers}
First, we examine whether $A_{AS}$ and $A_{DS}$ learn as the optimization objectives. Fig. \ref{fig:com} plots the sum of similarities for all speaker pair combinations per conversation, with each pair consisting of two different speakers, for the original data, $A_{AS}$, and $A_{DS}$ using predicted RTTM on clean simulation datasets. Overall, the sum of similarities for $A_{AS}$ is the lowest across all scenarios with different numbers of speakers, ensuring better speaker distinctiveness of anonymized speakers. $A_{DS}$ mimics the original conversation distribution, and the sum of similarity for each conversation matches that of the original conversation. However, as the number of speakers increases, it becomes more difficult to obtain a similar sum, resulting in more fluctuations and outliers.

\begin{figure*}[!t]
 \centering
 
 \subfloat[Resyn ]{\includegraphics[width=0.2\textwidth]{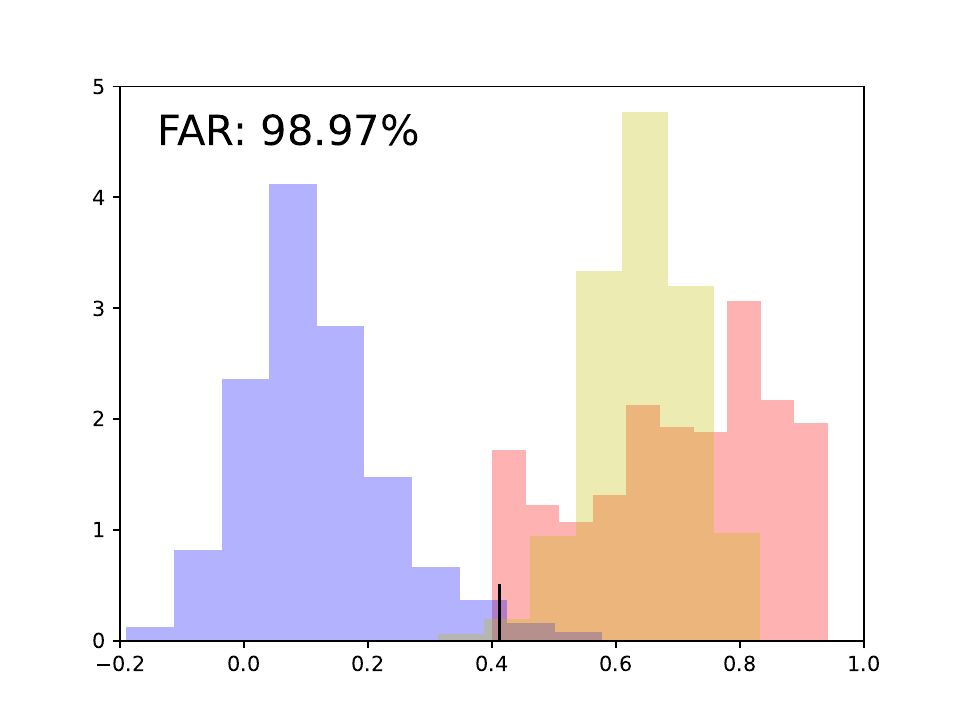}}
 \hfill
 \subfloat[$A_{OHNN}$ ]{\includegraphics[width=0.2\textwidth]{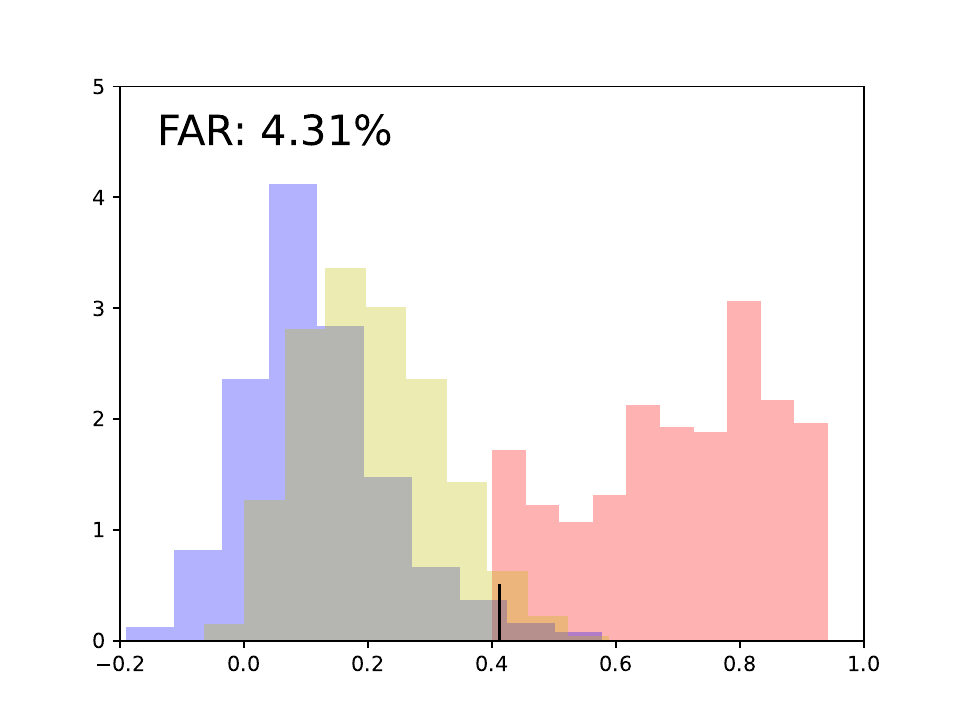}}
 \hfill
 \subfloat[$A_{Select}$ ]{\includegraphics[width=0.2\textwidth]{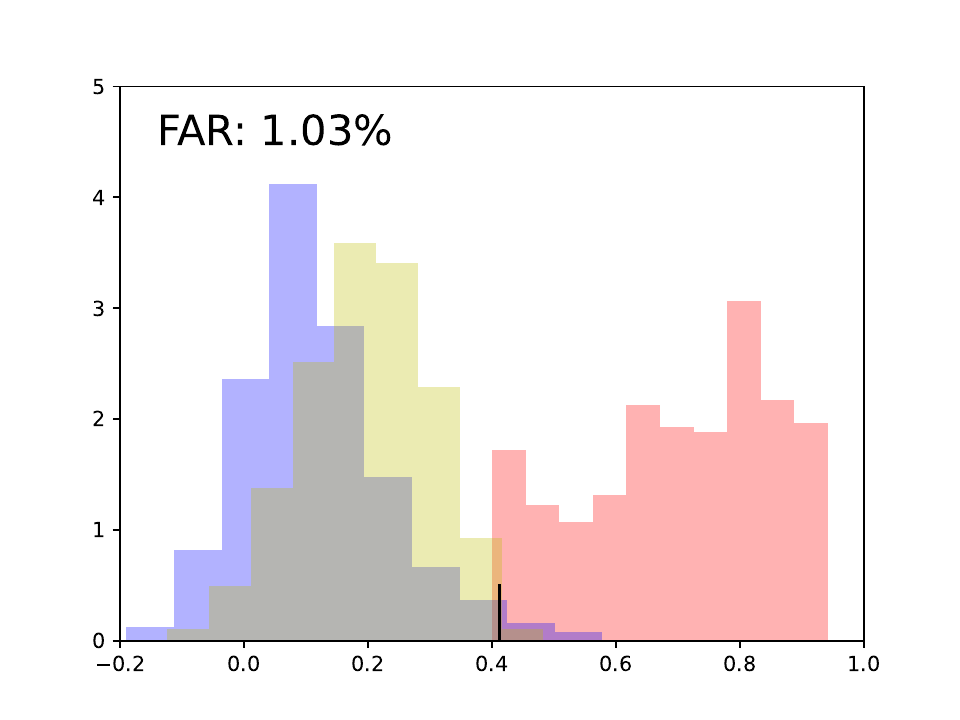}}
 \hfill
 \subfloat[$A_{DS}$ ]{\includegraphics[width=0.2\textwidth]{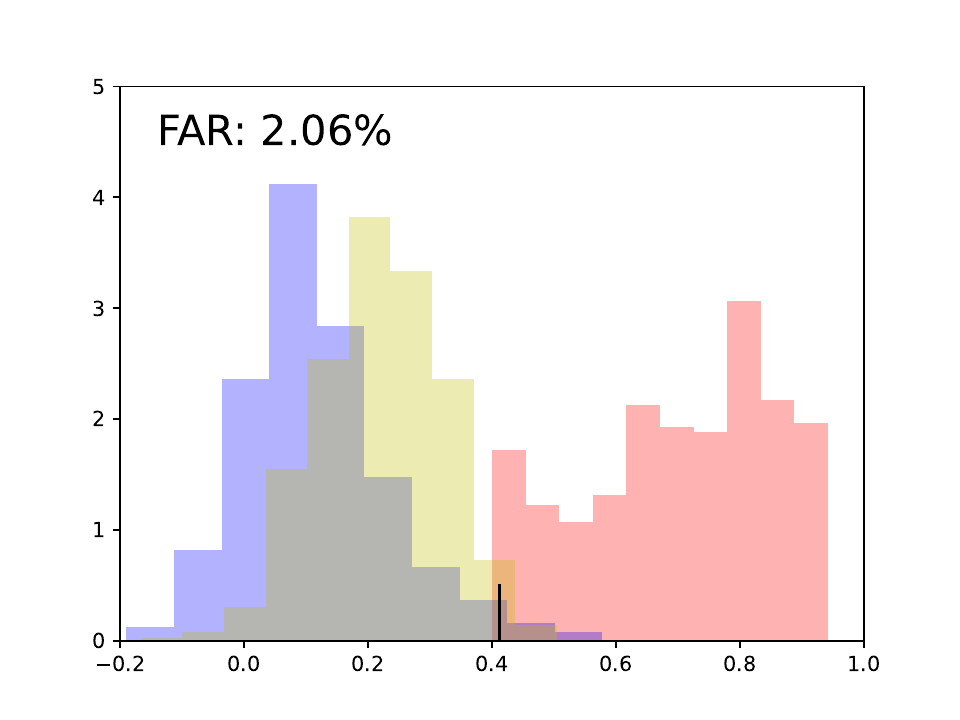}}
 \hfill
 \subfloat[$A_{AS}$]{\includegraphics[width=0.2\textwidth]{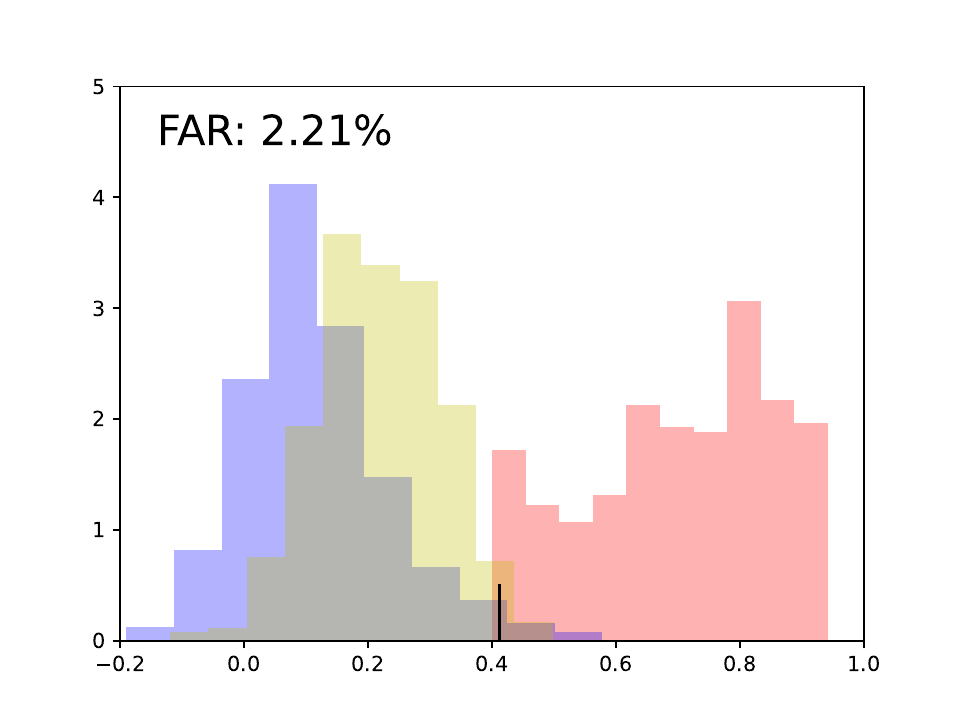}}
\vspace{-2mm}

\caption{Cosine similarities on 2-speaker clean conversation using predicted RTTM. The blue distribution represents the negative pairs formed by different speakers within one conversation. The red distribution represents the positive pairs formed by two 1.5-second segments split from 3-second single-speaker segments. The yellow distribution represents the pairs formed by original-anonymized segments.}

 \label{fig:predict-clean}
\end{figure*}

\begin{table}[]
 \centering
  \vspace{-2mm}
 \caption{WER(\%) $ \downarrow$ on original and anonymized audios}
 \begin{tabular}{c|ccccc}
 \toprule
 & Original & $A_{OHNN}$ & $A_{Select}$ & $A_{DS}$ & $A_{AS}$ \\
 \midrule
 Clean & 1.89 & 2.49 & 2.50 & 2.50 & 2.51 \\ 
 Noise & 5.58 & 13.44 & 13.69 & 13.70 & 13.70\\ 
 \bottomrule
 \end{tabular}
 \label{tab:single-spk}
 \vspace{-2mm}
\end{table}

\begin{table*}[]
 \centering
 \vspace{-2mm}
 \caption{DER(\%) $ \downarrow$ on original, resynthesized, and anonymized segments on simulated conversations using the real RTTM.}
 \begin{tabular}{c|c|ccccc|c|ccccc}
 \toprule
 & \multicolumn{6}{c|}{Clean} & \multicolumn{6}{c}{Noise} \\
 & Original & Resyn & $A_{OHNN}$ & $A_{Select}$ & $A_{DS}$ & $A_{AS}$ & Original & Resyn & $A_{OHNN}$ & $A_{Select}$ & $A_{DS}$ & $A_{AS}$ \\
 \midrule
 2-spk & 4.26 & 4.26 & 5.88 & 5.04 & 4.45 & \textbf{4.33} & 4.97 & 4.01 & 6.53 & 7.01 & 4.96 & \textbf{4.87} \\
 3-spk & 10.38 & 10.38 & 11.43 & 11.46 & \textbf{10.73} & 10.92 & 10.80 & 10.86 & 12.30 & 13.00 & 11.61 & \textbf{10.39} \\
 4-spk & 13.15 & 13.63 & 14.86 & 15.92 & 14.02 & \textbf{13.88} & 14.41 & 13.89 & 15.49 & 17.88 & 14.83 & \textbf{13.91} \\
 5-spk & 15.55 & 16.22 & 17.67 & 18.54 & 15.43 & \textbf{14.90} & 16.90 & 16.12 & 17.84 & 21.86 & 16.36 & \textbf{16.18} \\

 \bottomrule
 \end{tabular}
  \vspace{-2mm}

 \label{tab:real-rttm}
\end{table*}

\begin{table*}[]
 \centering
  \vspace{-2mm}
 \caption{DER(\%) $\downarrow$ on original, resynthesized, and anonymized segments on simulated conversations using the predicted RTTM.}
 \begin{tabular}{c|c|ccccc|c|ccccc}
 \toprule
 & \multicolumn{6}{c|}{Clean} & \multicolumn{6}{c}{Noise} \\
 & Original & Resyn & $A_{OHNN}$ & $A_{Select}$ & $A_{DS}$ & $A_{AS}$ & Original & Resyn & $A_{OHNN}$ & $A_{Select}$ & $A_{DS}$ & $A_{AS}$ \\
 \midrule
 2-spk & 4.26 & 5.51 & 14.82 & 7.06 & 5.92 & \textbf{5.86} & {4.97} & 6.00 & 15.82 & 7.69 & \textbf{6.11} & {6.43} \\
 3-spk & 10.38 & 11.30 & 19.82 & 13.20 & \textbf{11.61} & 11.91 & 10.80 & 11.88 & 22.37 & 13.33 & 12.22 & \textbf{11.79} \\
 4-spk & 13.15 & 14.67 & 27.04 & 17.08 & 14.90 & \textbf{14.74} & 14.41 & 15.17 & 28.89 & 18.16 & 15.59 & \textbf{15.91} \\
 5-spk & 15.55 & 16.63 & 33.17 & 19.98 & 17.46 & \textbf{17.15} & 16.90 & 17.98 & 34.45 & 21.12 & 18.64 & \textbf{17.69} \\
 \bottomrule

 \end{tabular}
  \vspace{-2mm}
 \label{tab:pre-rttm}
\end{table*}

\begin{figure*}[!t]
 \centering
 \subfloat[Real RTTM ]{\includegraphics[width=0.5\textwidth]{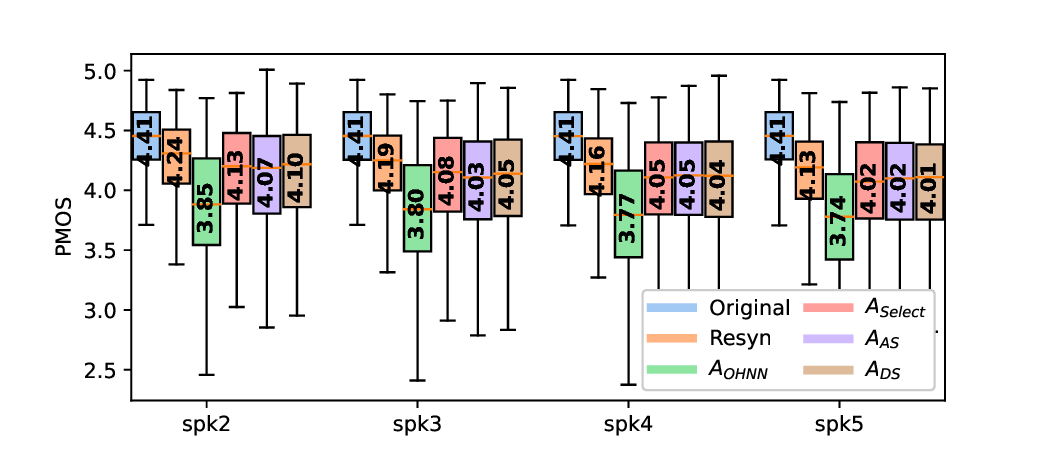}\label{fig:pmos-real}}
 \hfill
 \subfloat[Predict RTTM ]{\includegraphics[width=0.5\textwidth]{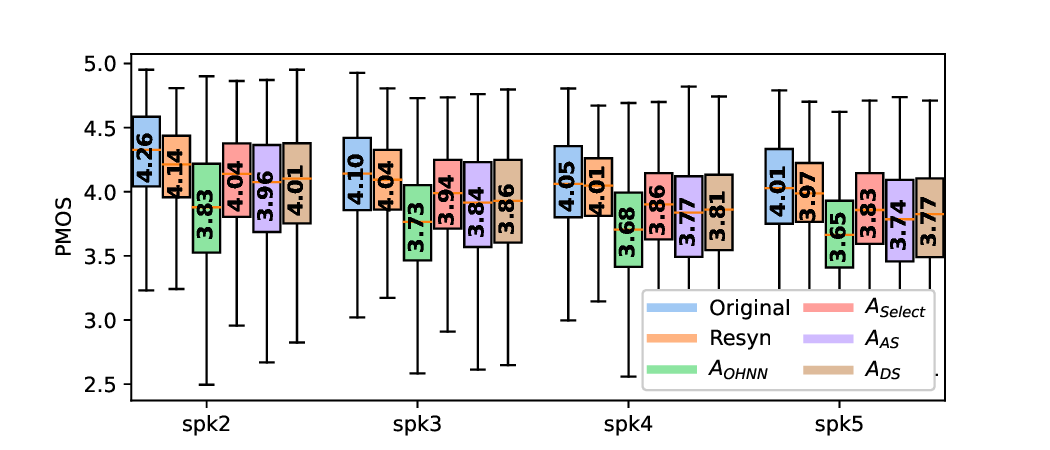}\label{fig:pmos-predict}}
 \hfill
 \caption{PMOS ($\uparrow$) on original, resynthesized, and anonymized segments split from clean simulation datasets.}
 \label{fig:pmos}
\vspace{-2mm}

\end{figure*}

\subsubsection{Results on simulated datasets}
\paragraph{Privacy Protection} Tables \ref{tab:real-rttm-far} and \ref{tab:pre-rttm-far} show the FARs for simulation datasets using real and predicted RTTM, respectively. Both tables exhibit similar trends: resynthesized speech without any protection/anonymization achieves nearly 100\% FAR, indicating complete leakage of the original speaker identity. With different MSAs, FARs can be reduced to less than 3\%, indicating effective speaker identity protection. To further visualize privacy protection ability, Fig. \ref{fig:predict-clean} plots the cosine similarities between pairs of speaker vectors extracted from original and anonymized segments for clean simulation datasets. The cosine similarity distributions of the original-anonymized pairs (yellow) have much less overlap with original-original pairs (red) in the right four subfigures, demonstrating the strong privacy protection offered by different anonymization methods.

\paragraph{Content preservation} The first row of Table \ref{tab:single-spk} shows the WERs on the original LibriSpeech test-clean subset used to simulate the conversation and its anonymized audios using different speaker vector anonymization methods. The second row presents the results for the noisy dataset, which adds background noises to the original LibriSpeech test-clean subset. All anonymized audios obtain higher WERs than those of original audios. For the clean condition, the absolute difference between the original (1.89\%) and different anonymized audios (around 2.50\%) is about 0.61\%. For the noisy condition, this difference increases to about $8.12\% = 13.70\% - 5.58\%$, as $ASR_\text{eval}$ is trained on clean speech, leading to more mismatches when decoding noisy speech. The differences among different anonymization methods are minor and the speech content is preserved at an acceptable level after anonymization.

\paragraph{Speaker distinctiveness preservation} Tables \ref{tab:real-rttm} and \ref{tab:pre-rttm} show the DERs for simulation datasets using real and predicted RTTM, respectively. Common conclusions from these tables include the following. (i) Resynthesized speech achieves DERs very similar to those of the original conversation, having almost no speaker distinctiveness loss. (ii) Noisy conversations achieve higher DERs compared with the same condition of clean conversations. (iii) As expected, a general progression in performance exists: $A_{DS}$ and $A_{AS}$ optimize the similarities among each speaker within one conversation, yielding lower DERs than $A_{Select}$ and $A_{OHNN}$, which are designed for SSA that neglects such similarities when anonymizing speech. (iv) In general, $A_{AS}$ achieves the best DERs among different anonymization methods, nearly similar to or slightly better than the corresponding original conversation, as it maximizes the similarities among each anonymized speaker, thereby obtaining better speaker distinctiveness after anonymization.

However, a few differences are observed. (i) Compared with real RTTM, using predicted RTTM increases the DERs overall because the segments for each speaker predicted by the SD system have errors where several frames are split to the wrong speaker. These cascading errors in turn affect DERs and typically introduce speaker confusion errors. (ii) The OHNN system achieves much worse DERs when using predicted RTTM. One potential reason is that the OHNN anonymizer is NN-based and very sensitive to input frames; even slight differences in input frames may result in anonymized speaker vectors belonging to entirely different speakers, leading to speaker confusion errors and increased DERs. Conversely, $A_{Select}$, $A_{DS}$, and $A_{AS}$ anonymized vectors on the basis of different similarity criteria, making them more robust to slight differences in RTTM.

\paragraph{Naturalness preservation} Fig. \ref{fig:pmos} plots the PMOS on original, resynthesized, and anonymized segments split from clean simulation datasets using real RTTM (right) and predicted RTTM (left). The first observation is that compared with using real RTTM, using predicted/inaccurate RTTM, which introduces discontinuous speaker segments, slightly decreases the overall naturalness. Additionally, there is a general trend in performance: the original speech performs best, followed by resynthesized speech. Next are $A_{Select}$, $A_{AS}$, and $A_{DS}$, which use different speaker vector selection strategies based on the same external pool, achieving similar PMOS and performing better than the $A_{OHNN}$.

In summary, the proposed MSA systems using $A_{{DS}}$ and $A_{{AS}}$ consistently achieve the best performance in terms of DER. For WER and PMOS, $A_{Select}$, $A_{{AS}}$, and $A_{{DS}}$ are comparable, with most cases showing slightly worse performance for $A_{{AS}}$ and $A_{{DS}}$ compared with $A_{Select}$. For FAR, when using real RTTM under clean conditions, $A_{Select}$ achieves the lowest FAR, and the differences between $A_{Select}$ and $A_{{DS}}$, $A_{{AS}}$ are the largest. However, these differences become progressively smaller under noisy conditions when using predicted RTTM, or with a larger number of speakers. This indicates that $A_{{AS}}$ and $A_{{DS}}$ are more robust.

\begin{table}[]
 \centering
  \vspace{-2mm}
 \caption{FARs (\%) on the VoxConverse dataset}
 \resizebox{\textwidth}{!}{

 \begin{tabular}{c|ccccc}
 \toprule
 RTTM & Resyn & $A_{OHNN}$ & $A_{Select}$ & $A_{DS}$ & $A_{AS}$ \\
 \midrule
 Real & 99.77 & 0.58 & 1.75& 0.70 & 0.00 \\ 
 Predict & 99.16 & 2.51 & 6.23 & 0.38 & 0.15 \\ 
 \bottomrule
 \end{tabular} }
 \label{tab:vox_far}
  \vspace{-2mm}
\end{table}

\begin{table}[]
 \centering
  \vspace{-2mm}
 \caption{DERs (\%) on the VoxConverse dataset}
 \resizebox{\textwidth}{!}{
 \begin{tabular}{c|c|ccccc}
 \toprule
 RTTM & Original & Resyn & $A_{OHNN}$ & $A_{Select}$ & $A_{DS}$ & $A_{AS}$ \\
 \midrule
 Real & \multirow{2}{*}{10.00} & 12.65 & 14.50 & 15.21& 14.48 & 13.75 \\ 
 Predict & & 13.34 & 15.51 & 16.54 & 15.25 & 13.72 \\ 
 \bottomrule
 \end{tabular}}
  \vspace{-2mm}
 \label{tab:vox}
\end{table}

\begin{figure*}[htbp]
 \centering
 \subfloat[ ]{\includegraphics[width=0.3\textwidth]{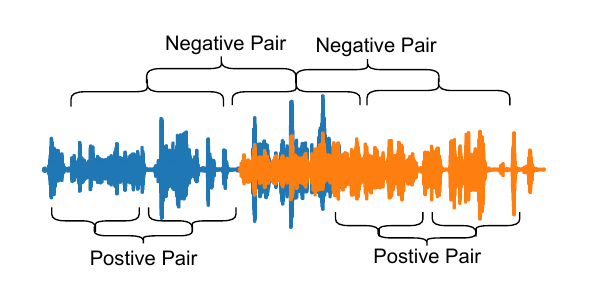}\label{fig:overlap-pairs}}
 \hfill
 \subfloat[ ]{\includegraphics[width=0.23\textwidth]{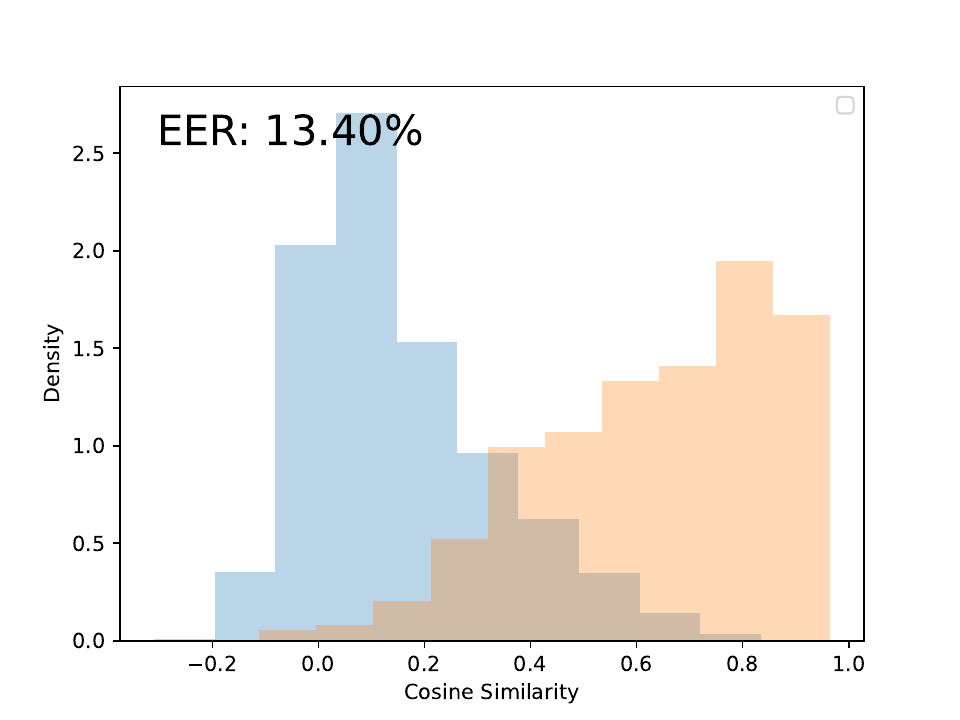}\label{fig:overlap}}
 \hfill
 \subfloat[ ]{\includegraphics[width=0.23\textwidth]{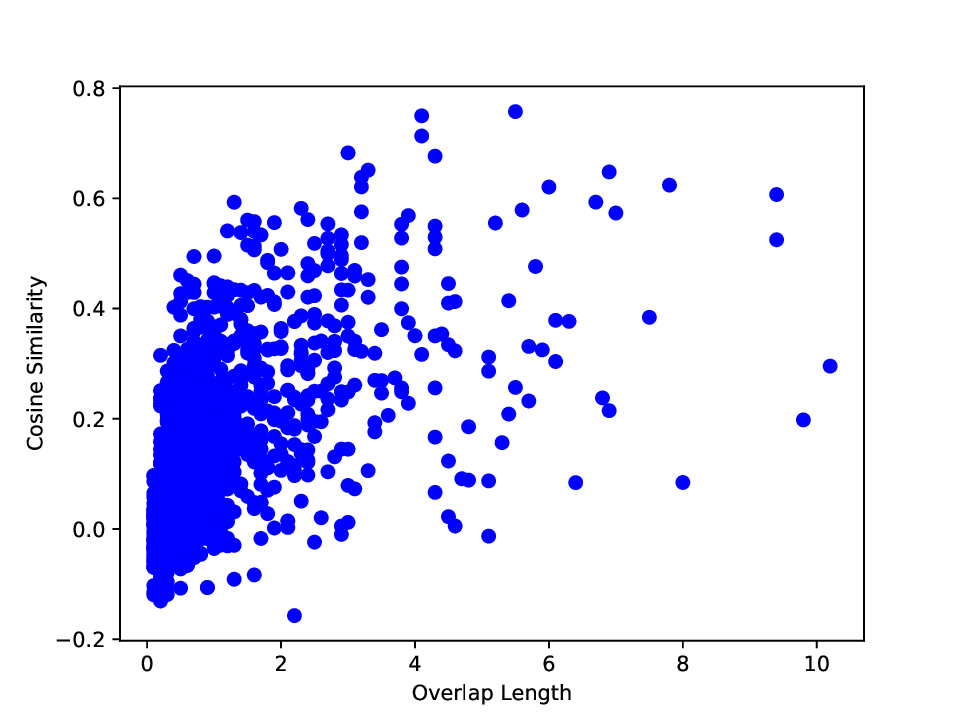}\label{fig:length-sim}}
 \hfill
 \subfloat[]{\includegraphics[width=0.23\textwidth]{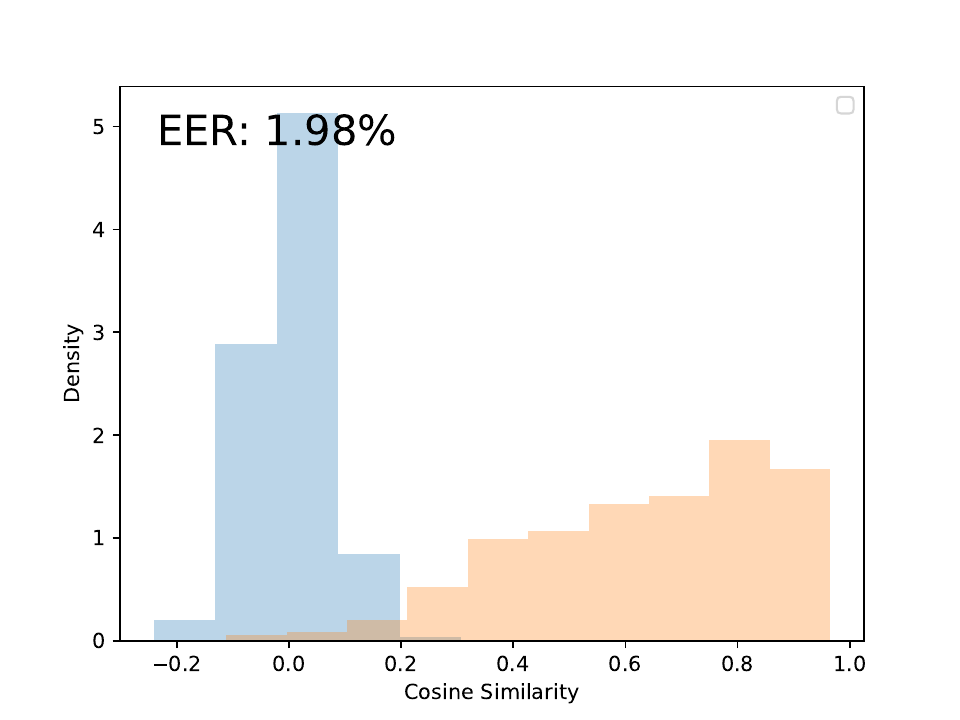}\label{fig:overlap-shuffle}}
\hfill
\vspace{-2mm}

\caption{Overlapping segments analysis. (a) Illustration of how negative pairs (overlapping and nearby single-speaker segments) and positive pairs (single-speaker segments split in half) are formed. (b) Cosine similarities of negative pairs (blue) and positive pairs (orange). (c) Relationship between the length of the overlap and the similarity of negative pairs. (d) Cosine similarities as in (a), but with overlapping segments shuffled along the timestamp.}
 \label{fig:overlap_ana}

\end{figure*}

\subsubsection{Results on VoxConverse datasets}
Tables \ref{tab:vox_far} and \ref{tab:vox} list the FARs and DERs for the VoxConverse test dataset, respectively\footnote{We omit WER and PMOS computations since VoxConverse dataset lacks a text transcript and includes real-world noise.}. The trends of FARs and DERs for both original and anonymized speech generated with different MSA systems are remarkably similar to those observed on the simulation test sets. Specifically, using $A_{DS}$ and $A_{AS}$ achieves less than 1\% FARs, showing almost perfect privacy protection ability. For DERs, $A_{AS}$ achieves about 13.7\% whether predicted or real RTTM is used, and is close to the original conversation, which achieved 10\%, showing good speaker distinctiveness preservation.

\section{Discussions}
\subsection{Discussion on overlapping speech}
\label{sec:overlap}
Until now, we have verified the effectiveness of proposed MSA systems on non-overlapping conversations. It is interesting to explore scenarios where overlapping segments exist, i.e., a cocktail party scenario, with a specific focus on the privacy risks to speakers in adjacent overlapping segments---both preceding and succeeding speakers. 

We consider the same attacker as in non-overlapping scenarios who uses $ASV_{eval}$ to infer the adjacent speaker identities from the anonymized overlapping segments\footnote{Note that a stronger attacker may infer the original speaker's identity after separating overlapping segments, we leave it as a future work.}. Specifically, we select 97 conversations from VoxConverse with overlaps and attempt to answer the following three questions.

\noindent
\textit{Does the overlapping segment reveal the privacy of nearby single speakers?} We crop overlapping segments from the 97 selected conversations and extract the connected preceding and following single-speaker segments. Overlapping segments and either preceding or following single-speaker segments are taken as negative pairs. The single-speaker segments, split in half, are taken as positive pairs, as shown in Fig. \ref{fig:overlap-pairs}.
Fig. \ref{fig:overlap} plots the cosine similarity between overlapping segments and their nearby single-speaker segments, including preceding and following single-speaker segments (blue), as well as the cosine similarity for segments from the same speaker (orange). We treated the former as negative pairs and the latter as positive pairs to compute the EER. Ideally, an EER of 0\% indicates the overlapping segments do not leak adjacent speaker identity information. However, an EER of 13.40\% was obtained, suggesting a low level of privacy leakage.

\noindent
\textit{Is the level of speaker privacy leakage related to the overlap length?} Fig. \ref{fig:length-sim} plots how the similarity of overlapping segments to nearby single-speaker segments changes with the length of the overlaps. There is a light trend that longer overlap lengths result in higher cosine similarity, indicating more speaker privacy leakage.

\noindent
\textit{What can be done to avoid privacy leakage from overlapping segments?} A naive approach is to sacrifice the utility when processing the overlapped segments. This can be done by detecting and shuffling the segments along the time axis. Fig. \ref{fig:overlap-shuffle} plots the cosine similarity, which is similar to Figure \ref{fig:overlap}, but with the overlapping segments shuffled along the timestamp. This shuffle reduces the EER from 13.40\% to 1.98\% as shown in Fig. \ref{fig:overlap-shuffle}. Despite the decreased EER, the naive method severely degrades the linguistic information in the overlapped segments. A potentially better pipeline may apply speech separation to the overlapped region, anonymize the separated segments, and reconstruct the utterance. This work raises awareness of the overlapping segments; however, the implementation and evaluation of more advanced pipelines are left for future work.

\subsection{Discussion on evaluation metrics}
In this paper, due to the high cost of recruiting human subjects for listening tests, we rely on objective metrics, including FAR, WER, PMOS, and DER. However, a key question arises: can objective evaluation metrics fully capture the subjective aspects of speech?
To assess the naturalness of anonymized speech, we use a MOS prediction network to compute PMOS. Previous work \cite{miao2023language} demonstrated a similar ranking of speaker anonymization systems between human-rated MOS scores and those predicted by the MOS network, suggesting that the network’s predictions align well with human perception. In addition, we use WER to evaluate intelligibility. Although the VoicePrivacy 2022 summary paper \cite{vpc2022sum} highlights that the correlation between WER and intelligibility is not particularly strong, it emphasizes that both objective and subjective metrics provide valuable insights.

\begin{figure}[t]
\centering
{\includegraphics[width=3in]{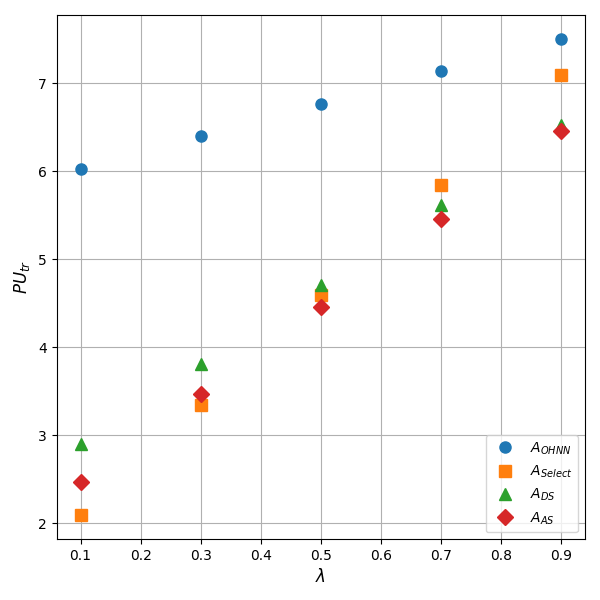}}
\caption{Privacy-to-utility trade-off for different MSA systems on clean simulated conversations, where FAR and DER are averaged across different numbers of speakers and calculated using the predicted RTTM.  Evaluations are conducted for $\lambda \in [0.1, 0.3, 0.5, 0.7, 0.9]$. }
\label{fig:single-metrix}
\end{figure}

Another limitation of the current metric computation is that the use of multiple evaluation metrics complicates the assessment process and makes comparison difficult, particularly because of the trade-off between privacy and utility, e.g. $A_{Select}$ achieves a lower FAR at the expense of conversational dynamics degradation, yielding a higher DER. To address this, we propose the privacy-to-utility trade-off, a compressed metric that combines the primary privacy metric and the primary utility scores, with a weight to control the importance of privacy and utility. Inspired by this, we adopt a similar compressed metric, privacy-to-utility trade-off ($\mathrm{PU_{tr}}$), which combines the primary privacy metric (FAR) and the primary utility scores (WER, DER, PMOS), designed to evaluate the anonymization system at different operating points. Given $\mathrm{WER_i}, \mathrm{DER_i}, \mathrm{PMOS_i}, \mathrm{FAR_i}, i\in\{0,1\}$ denotes the metrics calculated on original ($i=0$) or anonymized ($i=1$) utterances. 
\[
\footnotesize
\begin{aligned}
\mathrm{PU_{tr}} = &\ {\lambda} \left( \frac{\log\left(1 + \frac{\mathrm{WER_1}}{\mathrm{WER_0}}\right)}{\log\left(1 + \frac{1}{\mathrm{WER_0}}\right)} + \frac{\log\left(1 + \frac{\mathrm{DER_1}}{\mathrm{DER_0}}\right)}{\log\left(1 + \frac{1}{\mathrm{DER_0}}\right)} - \frac{\log\left(1 + \frac{\mathrm{PMOS_1}}{\mathrm{PMOS_0}}\right)}{\log\left(1 + \frac{1}{\mathrm{PMOS_0}}\right)} \right) \\
&+ (1 - \lambda) \cdot \frac{\log\left(1 + \frac{\mathrm{FAR_1}}{\mathrm{FAR_0}}\right)}{\log\left(1 + \frac{1}{\mathrm{FAR_0}}\right)}
\end{aligned}
\]
where $\lambda\in[0,1]$ controls the trade-off between utility and privacy. A lower $\mathrm{PU_{tr}}$ indicates a better trade-off at a specific operational point $\lambda$.
Figure \ref{fig:single-metrix} shows the $PU_{tr}$ results for different MSA systems on clean simulated conversations, where FAR and DER are averaged across different numbers of speakers and calculated using the predicted RTTM. Lower $\lambda$ prioritize privacy, while higher $\lambda$ emphasize utility. It is clear that as $\lambda$ increases, MSA using $A_{AS}$ and $A_{DS}$ achieves a better trade-off than other MSA systems, particularly when $\lambda > 0.5$. Within $A_{AS}$ and $A_{DS}$, $A_{AS}$ shows a better trade-off. This further confirms the effectiveness of $A_{DS}$ and $A_{AS}$, particularly in utility-DER.

\subsection{Discussion on MSA computational efficiency}

We have evaluated the effectiveness of the proposed MSA in terms of privacy and utility performance. However, it is equally important to analyze the latency, computational demands, and practicality of the MSA for real-world, latency-sensitive applications. The system achieves an average Real-Time Factor (RTF) of 0.29, which represents the ratio of the total processing time to 5 hours of audio\footnote{Simulations were conducted in both clean and noisy audio environments, involving 2–5 speakers, each contributing 5 hours of audio. The total processing time for eight different simulation datasets ranged from 4,444 to 7,057 seconds.}, as tested on an NVIDIA A5000 GPU, indicating that the system operates approximately three times faster than real-time. In terms of computational demands, the multispeaker anonymization system comprises approximately 130 million (130M) parameters, distributed among its key components: speaker diarization (15M), the HiFi-GAN Generator (14M), the HuBERT-soft Encoder (95M), and the ECAPA-TDNN (6M). Generating 540 seconds of audio with these models requires approximately 40 GB of RAM, making it costly and challenging to scale or deploy on edge devices. One possible solution, as proposed recently, is to replace traditional non-causal, computationally intensive networks (e.g., HuBERT and HiFi-GAN) with lightweight convolutional neural network architectures, thereby achieving low latency. Further optimization strategies are left for future research.
One possible solution, as proposed recently by \cite{quamer2024end}, is to replace traditional non-causal, computationally intensive networks (e.g., HuBERT and HiFi-GAN) with lightweight convolutional neural network architectures, thereby achieving low latency. Further optimization strategies are left for future research.

\subsection{Discussion on SSA and MSA computational efficiency and privacy-utility trade-offs}
It is evident that SSA is less resource-intensive, as it does not require speaker diarization or conversation-level speaker anonymizers. By directly applying SSA to multi-speaker anonymization, privacy is simultaneously enhanced, as all speakers are converted into a single speaker, making it difficult to identify the source speakers. However, this approach completely erases multi-speaker relationships, significantly reducing utility. 
Overall, SSA and MSA present different trade-offs: while SSA offers a more computationally efficient solution with greater privacy, it sacrifices utility by removing speaker-specific interactions. Conversely, MSA maintains multi-speaker relationships, offering higher utility but requiring more computational resources. The choice between SSA and MSA depends on the specific application and the relative importance of privacy and utility.

\section{Conclusion}

This paper established a benchmark for MSA, providing a flexible solution for anonymizing speech from different speakers. We developed a cascaded MSA system that uses spectral-clustering-based SD to accurately segment speakers. Each segment is then anonymized individually using a disentanglement-based method before being concatenated to reconstruct the full conversation. Additionally, we enhanced the selection-based speaker anonymizer, a critical component of the disentanglement-based method, by proposing two \textit{conversation-level} selection strategies. These strategies generate anonymized speaker vectors that improve speaker distinctiveness while ensuring the unlinkability of original and pseudo-speaker identities, and maintaining the distinguishability of pseudo-speakers within a conversation. We confirmed the effectiveness of the proposed MSA systems on both simulated and real-world non-overlapping conversations with various numbers of speakers and background noises. Finally, we discussed the potential privacy leakage caused by overlapping segments and provided possible lightweight solutions.

Future work could focus on: (i) Developing an end-to-end real-time MSA approach for overlapping and non-overlapping conversations: This paper presents a multi-speaker anonymization benchmark using a cascaded system, with our current focus primarily on non-overlapping conversations. However, overlapping speech segments also have the potential to reveal source speaker information and pose unique challenges for anonymization. Future work could investigate strategies to handle overlapping conversations effectively, even with an uncertain number of speakers. Additionally, improving the system's efficiency in terms of latency and computational demands is critical to enabling real-time applications, making this another important direction for future research. (ii) Enhancing evaluation metrics for a more comprehensive analysis: Although the proposed metric $\mathrm{PU_{tr}}$ is a novel attempt to integrate multiple objective evaluation metrics into a single measure, it relies on a hyperparameter (the weight $\lambda$) to balance privacy and utility. Future work could focus on refining the evaluation framework to streamline the assessment process, potentially by incorporating both subjective (e.g., human perceptual studies) and objective metrics. This would provide a more comprehensive analysis of the naturalness and practical utility of anonymized speech, addressing the limitations of the current approach.

\bibliographystyle{IEEEtran}
\bibliography{ref}

\begin{IEEEbiography}[{\includegraphics[width=1in,height=1.25in,clip,keepaspectratio]{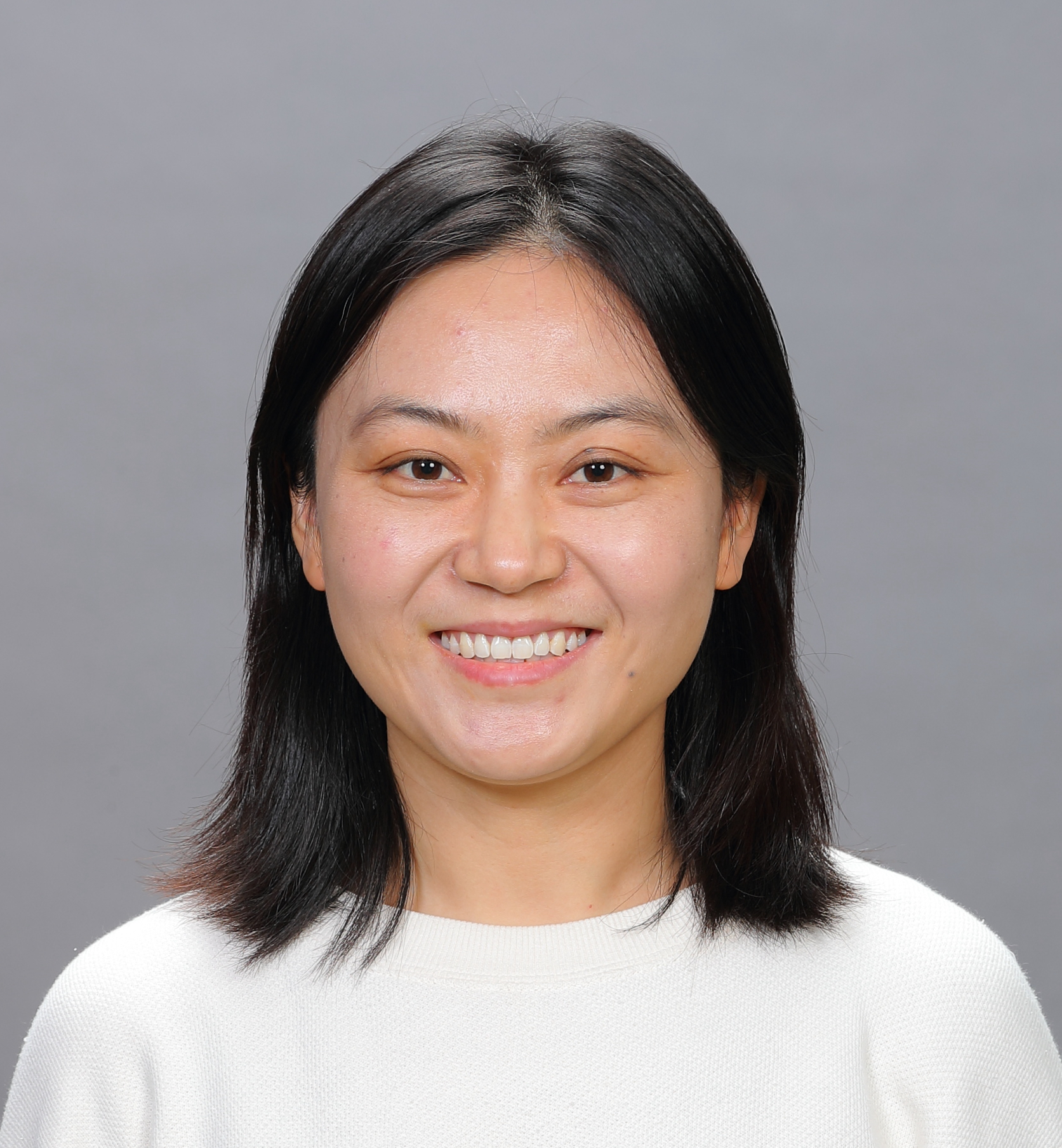}}]{Xiaoxiao Miao} (Member, IEEE) is an assistant professor at Singapore Institute of Technology. Prior to that, from 2021 to 2023, she was a postdoctoral researcher at the National Institute of Informatics (NII), Japan. She received the Ph.D. degree from the Institute of Acoustics, Chinese Academy of Sciences/University Chinese Academy of Sciences, in 2021. Her research interests include speaker and language recognition, speech security, and machine learning. She is a co-organizer of the latest VoicePrivacy challenge. 
\end{IEEEbiography}

\begin{IEEEbiography}[{\includegraphics[width=1in,height=1.25in,clip,keepaspectratio]{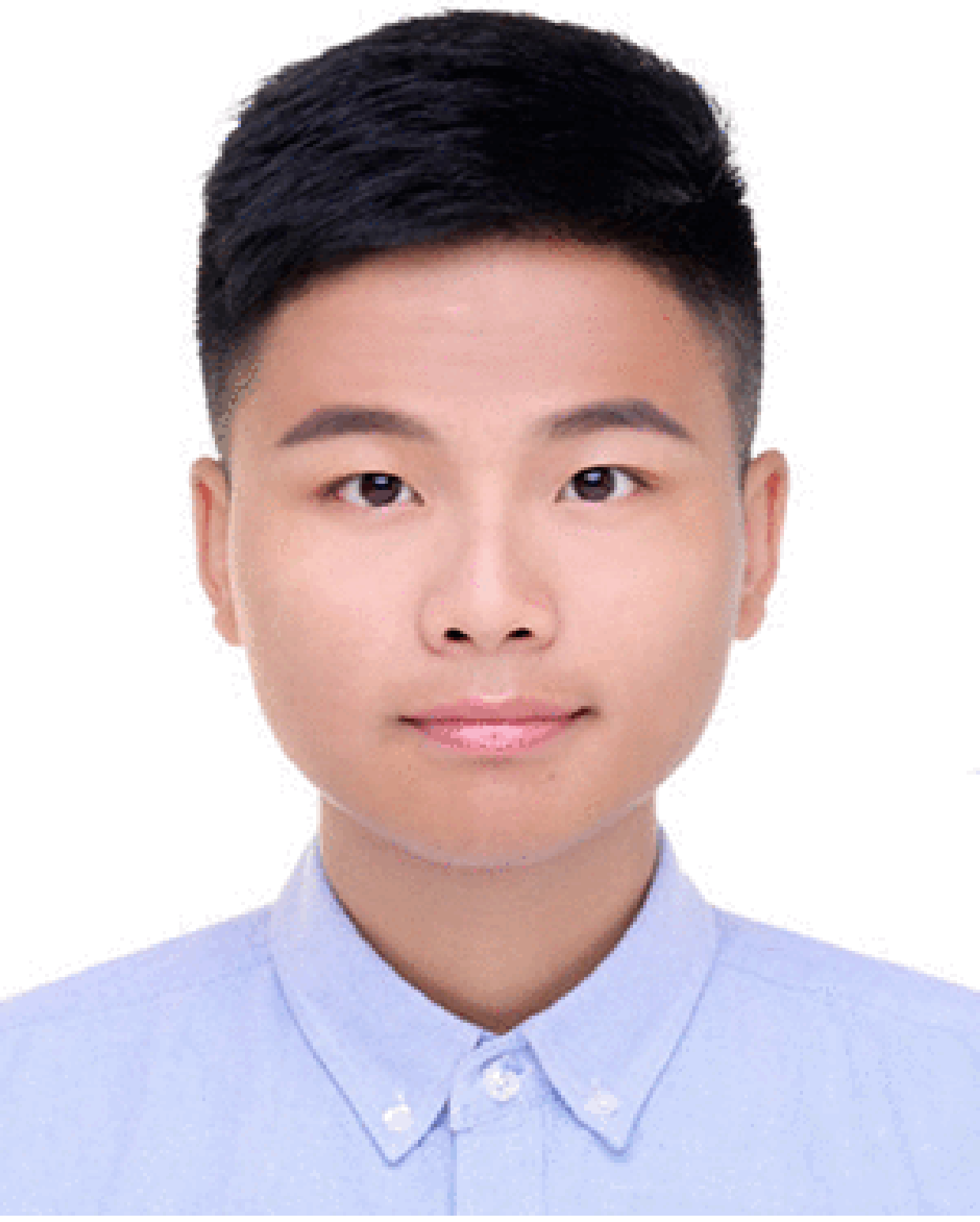}}]{Ruijie Tao} (Member, IEEE) received the Ph.D. and M.Sc. degree from National University of Singapore, Singapore, in 2023 and 2019, respectively. He received the B.Eng. degree from Soochow University, China, in 2018. He is currently a research fellow at National University of Singapore, Singapore. He is also the reviewer of CVPR, TASLP, ICASSP, Interspeech, SPL, CSL and SLT. His research interests include audio-visual speaker recognition, active speaker detection, speaker diarization, speech enhancement, speech extraction and anti-spoofing.
\end{IEEEbiography}

\begin{IEEEbiography}[{\includegraphics[width=1.15in,height=1.45in,clip,keepaspectratio]{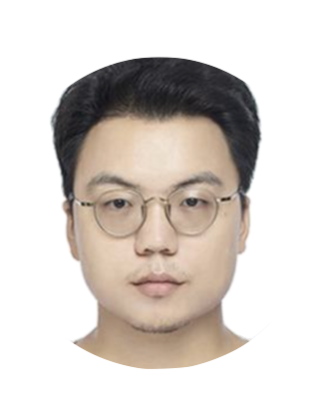}}]{Chang Zeng} received the Ph.D. from SOKENDAI, Japan, in 2024, and M.Sc. degree from The University of Tokyo, Japan, in 2020, respectively. He received the B.Eng. degree from Tianjin University, China, in 2016. He is also the reviewer of Neurips, ICLR, ICML, TASLP, ICASSP, Interspeech, and ICME. His research interests include LLM-based text-to-speech, audio generation, and neural audio codec.
\end{IEEEbiography}

\begin{IEEEbiography}[{\includegraphics[width=1in,height=1.25in,clip,keepaspectratio]{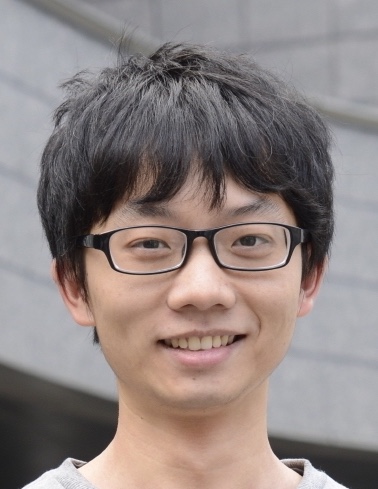}}]{Xin Wang} (Member, IEEE) is a project associate professor at the National Institute of Informatics (NII), Japan. He received the Ph.D. degree from SOKENDAI/NII, Japan, in 2018. Before that, he received M.S. and B.E degrees from the University of Science and Technology of China and University of Electronic Science and Technology of China in 2015 and 2012, respectively. His research interests include statistical speech synthesis, speech security, and machine learning. He is a co-organizer of the latest ASVspoof and VoicePrivacy challenges. 
\end{IEEEbiography}

\vfill

\end{document}